\documentclass{naturenew}
\usepackage{amsmath,amssymb}
\usepackage{graphicx}
\usepackage{epstopdf}
\usepackage{caption}
\usepackage{color}
\usepackage{amsmath}
 \usepackage{amssymb}
 \usepackage{graphicx}
 \newcommand{\be}{\begin{equation}}
 \newcommand{\ee}{\end{equation}}
 \newcommand{\br}{{\bf r}}
 
 \newcommand{\lf}{\left}
 \newcommand{\rg}{\right}
 \newcommand{\ra}{\rangle}
 \newcommand{\la}{\langle}
 \renewcommand{\a}{\alpha}
 \newcommand{\bea}{\begin{eqnarray}}
 \newcommand{\eea}{\end{eqnarray}}
 \renewcommand{\o}{\omega}

 \renewcommand{\d}{\delta}

 \title{No-go theorem for the description of Mott phenomena with
   conventional Density Functional Theory methods.}

 \author{Zu-Jian Ying$^{1,2}$\footnote{Present address: CNR-SPIN, and
     Dipartimento di Fisica ``E. R. Caianiello", Universit\`a
     di Salerno, I-84084, Fisciano (Salerno), Italy.}, Valentina
   Brosco$^1$, Giorgia Maria Lopez$^3$, Daniele
   Varsano$^4$, Paola Gori-Giorgi$^5$,  Jos\'e Lorenzana$^1$} 

\begin{document}

 \maketitle

 \begin{affiliations}
 \item ISC-CNR and Dipartimento di Fisica, Universit\`a di Roma 
 ``La Sapienza'', Piazzale  Aldo Moro 2, 00185 Roma, Italy.
 \item Beijing Computational Science Research Center, Beijing 100084, China.
 \item CNR-IOM, Istituto Officina dei Materiali, Cittadella
   Universitaria, Monserrato (CA) 09042, Italy
 \item Center S3, CNR Institute of Nanoscience, Via Campi 213/A, 41125 Modena, Italy.
 \item Department of Theoretical Chemistry and Amsterdam Center for
   Multiscale Modeling, FEW, Vrije Universiteit Amsterdam, Amsterdam
   1081 HV, Netherlands.
 \end{affiliations}

 \date{\today}

 \begin{abstract}
 Density functional theory provides the most widespread framework for
 the realistic description of the electronic structure of solids, but
 the description of strongly-correlated systems has remained so far
 elusive. Here we consider a particular limit of electrons in a periodic
 ionic potential in which a one-band description becomes exact all
 the way from the weakly-correlated metallic regime to the strongly-correlated Mott-Hubbard regime. We provide a necessary condition a
 density functional should fulfill to describe Mott-Hubbard behaviour and
 show that it is not satisfied by standard and widely used local,
 semilocal and hybrid functionals. We illustrate the condition in the
 case of a few-atom system and provide an analytic approximation to the exact exchange-correlation potential based on a variational wave function which shows
 explicitly the correct behaviour providing a robust scheme to
 combine lattice and continuum methods.  
 \end{abstract}

 Density functional theory (DFT) plays a fundamental role to understand 
 matter around us\cite{Kohn1999Nobel}, however it relies on
 approximations that fail to describe phenomena encountered in solids
 which have open d, or f shells such as 
 Mott-Hubbard insulating behavior\cite{Mott1949Basis},
 and correlation-induced band narrowing close to the Mott phase\cite{Imada1998Metalinsulator} and in
 heavy-fermions\cite{Coleman2006}. 
 It is commonly believed that these deficiencies
 ``cannot be remedied by  using more complicated exchange-correlation
 functionals in DFT'' (see Ref.~\cite{Kotliar2006Electronic}).
 Instead, these effects have been successfully
 described  by means of lattice models,  
 such as the Hubbard model,\cite{hub63,gut63,kan63}
  employing a series of techniques which have evolved into modern 
 Dynamical Mean-Field theory (DMFT)\cite{Georges1996}.  This  led to intense
 efforts to combine lattice and DFT methods\cite{Anisimov1991,Kotliar2006Electronic,Bunemann2005,Wang2008,Ho2008,Yao2011}. 

 In this work, we study a particular limit of correlated electrons in a solid 
 where  a one-band description becomes exact. More precisely the continuum
 model of electrons in a periodic potential and a generalized one-band 
 Hubbard model are bound to provide quantitatively equivalent solutions. 
 In this limit we derive a necessary condition a functional should satisfy to
 be able to describe Mott-Hubbard behavior. We show that most
 functionals in use today in chemistry and physics, including local,  
 semilocal and hybrid functionals,\cite{martinbook}
  do not satisfy the condition and they cannot
 describe Mott-Hubbard phenomena. This provides a
 rigorous ground to the quoted statement of Ref.~\cite{Kotliar2006Electronic}  when restrained to conventional approximations.  On the
 other hand,  we also show that
 in simple test cases a suitable combination of DFT and lattice methods can
 produce an exchange-correlation (xc) potential,  that not only shows the correct qualitative behavior, but it is also in surprisingly good
 agreement with results in test systems tractable with accurate
 wave-function methods.

 \section{Results}

 \subsection{Continuum model and one-band limit}
  We consider a fictitious system consisting of $N$  electrons moving in the
 potential $v_{\rm ext}$ produced by a collection of $N$
 identical nuclei of charge $Z e$ located on the sites ${\bf  R}_i$
 of a periodic lattice with spacing $a$. The system can be viewed as a
 half-filled 1s band in which $Z$ controls the
 orbital size, $a_{\rm B}/Z$, with  $a_{\rm B}$ the Bohr
 radius. Using  $a_{\rm B}/Z$ as the unit of length and  $Z^2$Ha as the unit of energy the Hamiltonian reads, 
 \begin{equation}
     \label{eq:hcont}
 H= \sum_i^{N}\hat h({\bf
   r}_i)+\frac{1}{2Z}\sum_{i\ne j}^{N}\frac{1}{|\br_i-\br_j|}
 \end{equation}
 where   ${\bf r}_i$ are the electron coordinates  and $\hat h({\bf r})$ is the one-body Hamiltonian, 
 $\hat h({\bf r})=-\frac12 \nabla_{{\bf r}}^2+v_{\rm ext}({\bf r})$ with,
 $$v_{\rm ext}({\bf r})=-\sum_{j}^{N}\frac{1}{|\br-{\bf R}_j|}.$$
 Notice that, in  these rescaled units, the one-body part becomes
 $Z$-independent and $1/Z$ plays the role of a coupling
 constant\cite{Ullrich2001}. We will use $a$ and $Z$ as our control variables.

   Our first goal is to find a region of the parameter space ($a,Z$), where a
   one-band (1B) lattice model describes quantitatively and not only qualitatively the continuum model defined by equation \eqref{eq:hcont}. 
 As it is well known from studies of the Hubbard
 model,\cite{hub63,kan63,gut63,bri70,geo96} the
 most important energy scales in this problem are the Hubbard 
 on-site interaction, $U$, and the nearest-neighbor hopping matrix
 element, $t$, which define the weakly-correlated regime, $0<U\ll z t$, and the
 strongly-correlated regime, $U\gg z t$, $z$ denoting the lattice
 coordination.  

\begin{figure}
 $$\includegraphics[height=5cm]{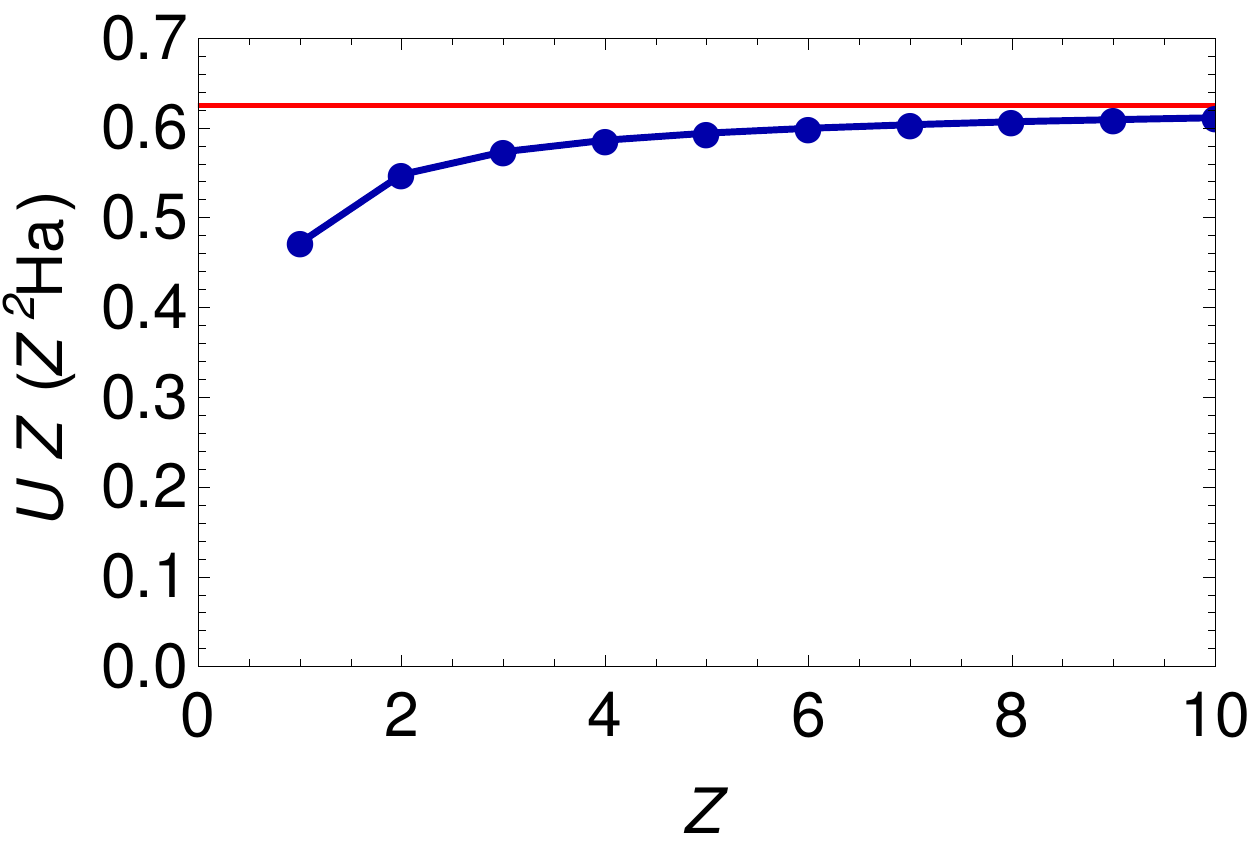}$$
 \caption{{\bf Vanishing of screening effects at large atomic number.}
   The 1s Hubbard $U$  vs. the atomic
   number. We multiply $U$ by $Z$ so that the limiting value is a
   constant in our units.  
 The red line is the bare value $U_0$ from Coulomb integrals [equation~\eqref{eq:coulint}],
  while the blue dots are the ``screened'' value obtained as
 $U=E_I^{Z,Z}-E_I^{Z,Z-1}$ where $E_I^{Z,N}$ is the $N$-th ionization
 energy for atomic number Z. 
 ($E_I^{1,0}\equiv E_A$ is the affinity energy for
 Hydrogen)\cite{Kramida2012}.
 }
 \label{fig:uzdz}  
 \end{figure}

 In the atomic limit,
 for $Z=1$, the elimination of higher energy bands is inaccurate, since
 the energy cost\cite{Kramida2012} $U=0.47 $ Ha of a charge fluctuation  $1s^1
 1s^1 \rightarrow 1s^0 1s^2$ is similar to the cost $\Delta=
 3/8$~Ha  of a  charge fluctuation $1s^1 1s^1\rightarrow
 1s^1 2s^1$.   
 To avoid this problem we
 take the limit of large $Z$ adjusting $a\sim \log(Z)$ so that $t/U$ is kept constant (hereafter ``the 1B limit''). In this limit  the condition
 \begin{equation}
   \label{eq:largedel}
 \Delta\gg t,U, %...
 \end{equation}
 is fulfilled allowing us to study  quantitatively  Mott-Hubbard phenomena in
 the model equation~\eqref{eq:hcont}  from  the weak to the strongly-correlated
 regime within a 1B description.  Figure~\ref{fig:uzdz} shows how this
 works for the Hubbard repulsion. As $Z$ grows the ``screened'' $U$ 
 asymptotically approaches the bare $U_0$ 
  showing that corrections from higher orbitals become irrelevant.

 \subsection{Lattice Model.}

 In the 1B limit the continuum model is equivalent to a generalized
 Hubbard model which can be obtained by the standard
 second-quantization procedure employing a single particle basis of 1s
 orbitals $\phi_i({\bf r})$  centered at ${\bf R}_i$ (see Methods and
 Refs.~\cite{hub63,kan63,gut63,Amadon1996,Spaek2000,Brosco2015}). 
  It will become clear below that our arguments are sufficiently
  general that are valid for such generalized Hubbard model. 
 However to fix ideas it is useful to think in terms of the standard
 Hubbard model\cite{hub63,kan63,gut63} which can be obtained by 
 absorbing the long-range part of the Coulomb interaction in mean-field 
 in the on-site energy $v$ and restricting the hopping 
 to nearest neighbors $\langle ij \rangle  $, 
 \begin{equation}
 H_H=\sum_{i\sigma}v\, n_{i\sigma}-t\sum_{\langle ij \rangle  \sigma}(c_{i\sigma }^{\dag}c_{j\sigma
 }+{\rm H.c.})+U\sum_i n_{i\uparrow }n_{i\downarrow },\label{eq:hub}
 \end{equation}
 with $c_{i\sigma}$  ( $c^{\dag}_{i\sigma}$)
 the annihilation  (creation) operator for an electron with spin
 $\sigma$ on site $i$ and $n_{i\sigma}\equiv c^{\dag}_{i\sigma}c_{i\sigma}$.

  To  link  lattice and DFT descriptions, we
 need the correlated density  which is determined
 by the one-body density matrix of the lattice model ($\rho_{ij}\equiv\sum_\sigma
 \langle c_{j\sigma}^\dagger  c_{i\sigma} \rangle$) and the $\phi_i$'s (Methods). The density can be separated into an ``atomic'' component  and a ``bond-charge'' contribution, $n({\bf  r})=n_{\rm at}({\bf r})+n_{\rm
   bd}({\bf r})$ with,   
  \begin{eqnarray}
  n_{\rm at}({\bf r})&=&\sum_{i}  |\phi_i({\bf r})|^2\rho_{ii}  \label{eq:natom}\\
 n_{\rm bd}({\bf r})&=&\sum_{\langle ij \rangle} \phi_j^*({\bf r})\phi_i({\bf r}) \rho_{ab}+c.c. ,\label{eq:bondcharge}
  \end{eqnarray}
 where, we used the fact that for large $a$ the sums in equation~\eqref{eq:bondcharge}
 can be restricted to nearest-neighbors and we defined the nearest-neighbor density-matrix element $\rho_{ij}=\rho_{ab}$.
 In our case, $\rho_{ii}=1$, thus all the dependence of the density on
 $U/t$ is encoded in the bond charge and it is controlled by $\rho_{ab}$. 
 In the limit of large $Z$, we can study the system from small to large
 $U/t$ either varying $a$ or changing $Z$ at a fixed
 $a$. In the latter case, since the $\phi_j$'s can
 be taken as fixed independently of the value of $U/t$, 
 equation~\eqref{eq:bondcharge} establishes a one to one mapping between the
 density and  $\rho_{ab}$ as  $U/t$ (or equivalently $Z$) is varied.

 \subsection{Hartree approximation}

 Before examining DFT, we analyze the ground state of Hamiltonian
 equation~\eqref{eq:hcont} in the Hartree approximation. As an initial guess
 of the Hartree orbitals, we can take the eigenstates of the
 non-interacting problem.  We will show below that the Hartree potential $v_H({\bf r})$ is of
 order $O(1/Z)$, thus, since $\Delta$ is $O(Z^0)$, the corrections to the
 initial guess can be neglected and, to leading order in $1/Z$, the Hartree orbitals coincide with the non-interacting orbitals and the Hartree self-consistent
  density coincides with the non-interacting density. To fix ideas
  consider the case $N=2$, although the arguments are easily generalized
  to any $N$.  The initial guess of the occupied Hartree orbital is the bonding state,  
 $$
 \psi_0^{\rm H}({\bf r})=\frac1{\sqrt2}[\phi_a({\bf r})+\phi_b({\bf  r})].$$ 
 Here $i=a,b$ labels the two ions and we assume real orbitals.
 The Hartree  density $n^H({\bf r})$ is given by $n^H({\bf r})=2
 |\psi_0^{\rm H}({\bf r})|^2 $ or equivalently  by eqs.~\eqref{eq:natom},\eqref{eq:bondcharge}
  inserting $\rho_{ij}$ in the Hartree approximation,
  $\rho_{ij}^H$, where for two-sites $\rho_{ab}^H=1$.  

 The Hartree potential is given by, 
 $$v_H({\bf r})=\frac1Z \int d^3{\bf r}' \frac{ n^{H}({\bf r}')}{|{\bf r}-{\bf
     r}'|}.$$ 
  This is finite everywhere and it has a maximum that scales at most as $1/Z$.
  Thus, we can always choose
 $Z$ large enough so that $v_H/\Delta\sim 1/Z$ is 
 small and the non-interacting orbitals coincide with the Hartree
 orbitals.

 \subsection{How Mott-Hubbard correlations are encoded in the density}

 \begin{figure}[t]
 \begin{center}
 \includegraphics[height=5cm]{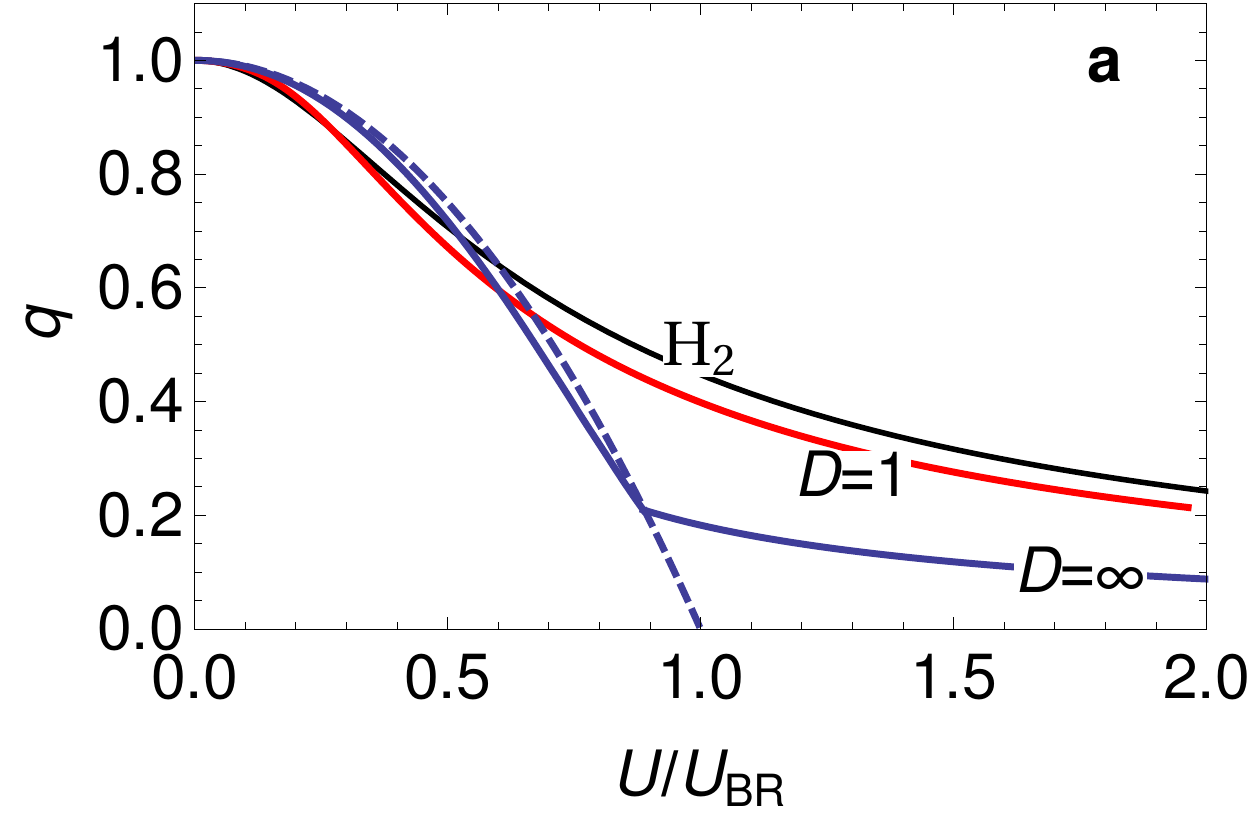}
 \includegraphics[height=5cm]{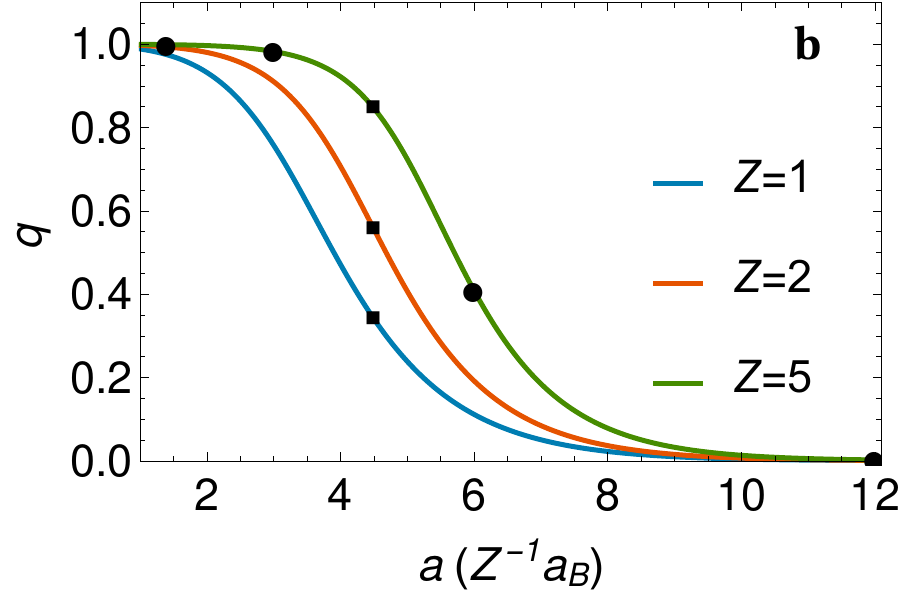}
 \caption{{\bf Hopping reduction factor.}
 {\bf a} Exact results (full lines) as a function
   of $U/U_{\rm BR}$ for the Hubbard model in dimension $D$.  Here $U_{\rm BR}$  is a measure of
   the effective non-interacting 
   bandwidth\cite{Vollhardt1984},  $U_{\rm BR}\equiv \frac{16}N \sum_{k\in {\rm
       occ.} } \epsilon_k$ where $\epsilon_k$ are the single particle
   non-interacting energies and the sum is restricted to the occupied
   states. We show results for a two-site system 
   (black),  for an infinite
   chain (red) and for the Bethe lattice in infinite dimensions
   (blue).  The Bethe lattice data
 where derived from numerical\cite{Karski2005Electron} and
 analytic\cite{Blumer2005Ground} results (full blue line).
 The cusp signals $U_{\rm c2}$ where the Mott-Hubbard transition occurs
 from the metallic phase ($U< U_{\rm c2}$) to the insulating phase ($U>
 U_{\rm c2}$).  The dashed line was obtained with the Gutzwiller wave
 function for the same Bethe lattice. {\bf b} Hopping reduction factor
 obtained with a Gutzwiller type wave function in an hydrogenic
 molecule as a function of atom separation $a$ for different atomic
 numbers. The dots (squares) correspond to the cases considered in
 Fig.~\ref{potz5}  (Fig.~\ref{fig:rhodr} and Supplementary Fig.~\ref{potr4p5}).}
 \label{fig:qdu}  
 \end{center}
 \end{figure}

 Since DFT provides an exact description\cite{Kohn1999Nobel}, Mott-Hubbard 
 behavior should be encoded in the density. 
  In the present limit this can only happen through the bond charge
  equation~\eqref{eq:bondcharge}  which can be parametrized through the
   hopping reduction factor defined as,  
 $$
 q\equiv\frac{\rho_{ab}}{\rho_{ab}^H}.
 $$
 For an interacting system, one  typically finds  $q<1$. For example  
 Fig.~\ref{fig:qdu}a shows the hopping reduction factor of the
 Hubbard model for exactly solvable lattices. In more general cases, a
 good estimate of the hopping reduction factor can be obtained with a variational  Gutzwiller
 wave function (see \ref{gwf}). 
 For the two-site Hubbard model, this yields the exact solution while for
 infinite dimension it provides an accurate estimate in the
 metallic phase (dashed line in Fig.~\ref{fig:qdu}).  

 For the full model equation~\eqref{eq:latticemodel}, the 
 qualitative behavior of the hopping reduction factor does not
 change as it can be easily checked using perturbation theory. 
 Fig.~\ref{fig:qdu}b shows the hopping reduction factor computed
 with a variational Gutzwiller-type wave function for the 
 two-site full model  (Methods)  as a function of the interatomic distance, $a$. 
  Since $t$ decreases exponentially with $a$ one obtains a rapid
  crossover from the  weakly-correlated regime ($q\sim 1$) to the strongly-correlated regime  ($q\sim 0$).
 It is believed that the crossover turns into the Mott  metal-insulator
  transition in high dimensional lattices as  Fig.~\ref{fig:qdu}a suggests.

\begin{figure}[tb]
 $$\includegraphics[width=8cm]{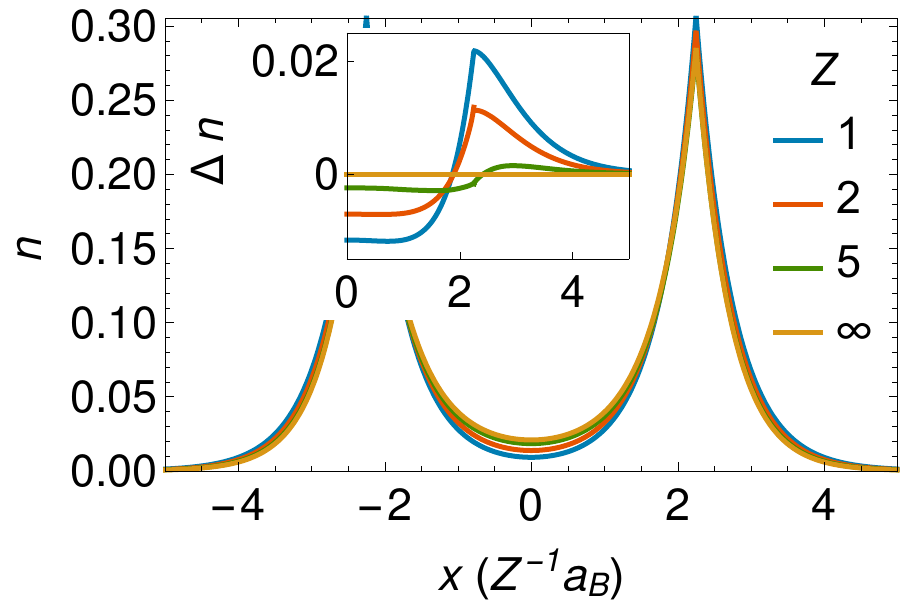}$$
 \caption{Charge along the bond for interatomic distance $a=4.5\  a_B Z^{-1}$ and various
 values of $Z$. The inset shows the interacting charge minus the
 nointeracting charge ($Z=\infty$). 
 Charges where computed within full CI (see Methods). 
 }\label{fig:rhodr}  
 \end{figure}

  From the above discussion it is clear that in the 1B limit, 
 independently of the dimensionality, the correlated charge in the bond
 is depleted with respect to a Hartree computation ({\it c.f.}
 equation~\eqref{eq:bondcharge} and Fig.~\ref{fig:qdu}). Figure~\ref{fig:rhodr}
 shows an accurate computation of the density within  the full 
 configuration interaction (CI) approach. 
 For small $Z$ (strong correlation) the bond charge is indeed depressed, 
 however, the differences are minute which makes
  extremely challenging for DFT to be sensitive to Mott-Hubbard correlations. 

 \subsection{No-go theorem.}
  In Kohn-Sham (KS) theory\cite{kohn1965} the interacting system is mapped
  into a system of  non-interacting electrons moving in an effective
  potential which is the sum of the external potential, the Hartree (H)
  potential and the xc-potential, 
 \begin{equation}
   \label{eq:vks}
 v_{\rm KS}({\bf r})=v_{\rm ext}({\bf r})+v_{\rm H} ({\bf r})+ v_{\rm xc} ({\bf r}).
 \end{equation}
 In the exact formulation the non-interacting system  reproduces the
 exact density of the interacting system, however in 
 practice $v_{\rm xc}$ is an unknown functional of the density
 and approximations are needed. 
 In the local density approximation\cite{kohn1965,martinbook}
 (LDA), the xc-potential is a simple function of the local density,
 $v^{\rm LDA}_{\rm xc} ({\bf r})=v_{\rm xc}^{\rm LDA}[n({\bf r})]$. We show now that LDA
 cannot account for Mott-Hubbard correlations of the system.

 To solve KS equations in  LDA one  can use again the Hartree density,
 $n^H$, as a starting guess. Using a standard parametrization of the
 potential, it is easy to show that $v^{\rm LDA}_{\rm Hxc}$ is at most of order
 $\sim 1/Z$ (see \ref{sec:testing-functionals}). For large $Z$, the change in the orbitals is then of order
 $v_{\rm xc}^{\rm LDA}/\Delta\sim 1/Z$ and it can be neglected. Thus in the 1B limit LDA orbitals coincide with the Hartree
 orbitals and the density is given by
 equations~\eqref{eq:natom},~\eqref{eq:bondcharge} with $\rho_{ij}^{\rm LDA}=\rho_{ij}^H$ independently of $U/t$.
 It is clear that LDA cannot account for the bond-charge 
 reduction which is a primary characteristic of Mott-Hubbard
 correlations.

 The  failure of LDA can be traced back to the  $1/Z$ scaling of $v_{\rm
   xc}$. Clearly, the 
 exact  $v_{\rm xc}$ cannot scale as  $1/Z$ everywhere, since if it did so,  the same argument as for  LDA  would apply. The only way to
 generate  the correlated density with a non-interacting system is 
 by modifying the orbitals, and this can only happen by allowing the
 low-energy 1s states to be mixed with high energy single particle
 states outside the minimal basis set, consistently with the known fact
 that exact KS-DFT breaks down when restricted to a finite basis set.\cite{Schipper1998,SavColPol-IJQC-03}

  For large $Z$ the mixing of the 1s band with the higher bands 
 can only be achieved if the xc-potential is of order $Z^0$. It
 follows that a necessary condition that a functional must satisfy to
 describe Mott-Hubbard behavior in the limit in which the 1B mapping 
 is accurate is that the potential should have regions which scale as
 $Z^0$.  Any functional whose xc-potential scales to zero when
 $Z\rightarrow \infty$ (keeping $U/t$ constant) cannot describe
 Mott-Hubbard behavior. This is the case for local semilocal
 and hybrid functionals\cite{Leeuwen1994} (see \ref{sec:testing-functionals}).
 and even the exact DFT strongly-correlated
 limit\cite{Gori-Giorgi2010,MalMirGieWagGor-PCCP-14}, which uses a
 wavefunction with maximally correlated electrons to compute the
 expectation of the electron-electron interaction, and becomes exact in the
 limit $a\gg Z$, but not in the present case, $a\sim\log(Z)$.

 \subsection{Mott barriers}

 How does the exact xc-potential look like? It should have a barrier in
 the bond to deplete the KS density, a well-known 
 result from the study of molecules.\cite{Leeuwen1995Step,Gritsenko1996Molecular,Leeuwen1996,Gritsenko2000Comparison,Helbig2009Exact}
 We term the part of the xc-potential that scales as
 $Z^0$ a ``Mott barrier''. 
 It can be proved that the barrier height in the strongly-correlated
 regime is related to the ionization potential of the 
 system;\cite{Leeuwen1995Step,Gritsenko1996Molecular,Leeuwen1996,Gritsenko2000Comparison,Helbig2009Exact}
 it is therefore of order $O(Z^0)$ as expected from the previous
 arguments.

 \begin{figure}
 $$\includegraphics[height=5cm]{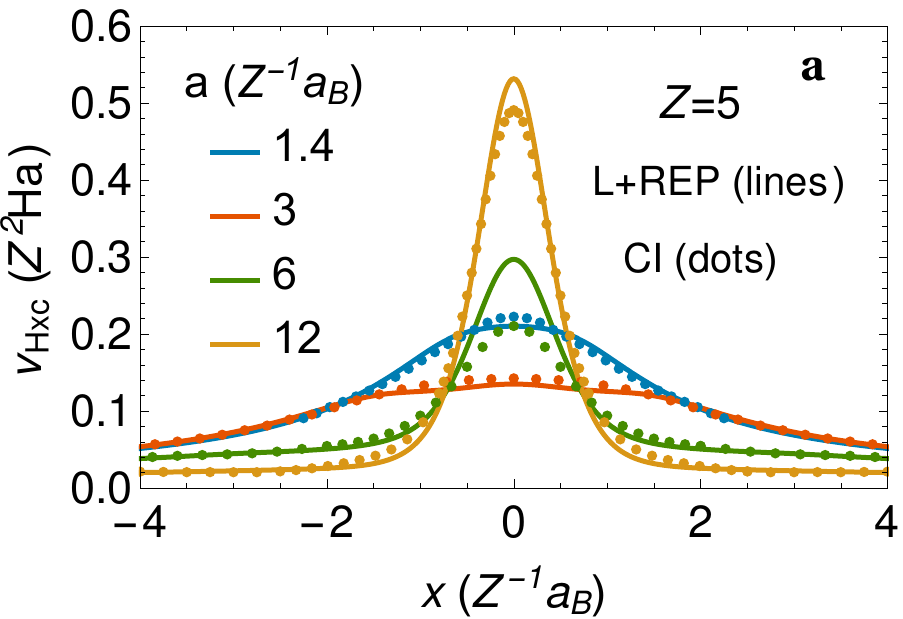}
 \includegraphics[height=5cm]{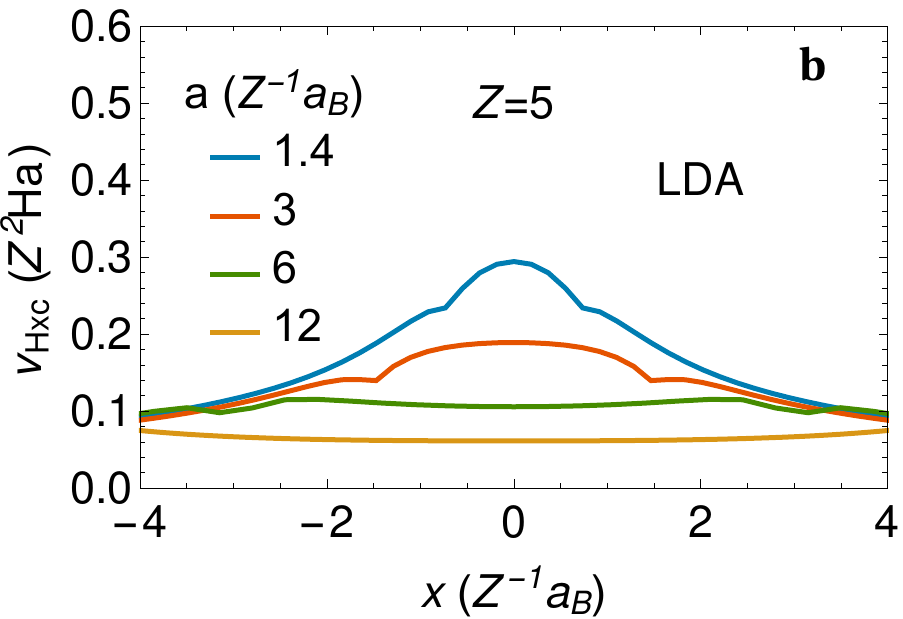}$$
 \caption{{\bf Hartree-exchange-correlation potential for a two-site
     system.} The potential is plotted along the bond direction $x$ for
   $z=y=0$,  $Z=5$ and different values of the interatomic distance
   $a$.  The dots in {\bf  a} are obtained inverting the full  CI density
   while panel  {\bf  b} shows the 
   LDA.~\cite{cp2k}   The lines in
   {\bf a} are the L+REP approximation using orthogonalized atomic
   orbitals.  The
  strength of the correlations for each case is defined by the value of $q$ in Fig.~\ref{fig:qdu}.
 }
 \label{potz5}
  \end{figure}

 To illustrate the Mott-barrier effect and develop an approximation for the exact potential, we follow a long
 tradition\cite{Mott1949Basis,Anderson1950Antiferromagnetism} and study a two-site system as a proxy for the lattice.  As in the pioneering works of  Refs.~\cite{Leeuwen1995Step,Gritsenko1996Molecular,Leeuwen1996,Gritsenko2000Comparison,Helbig2009Exact}
 we compute the Hxc-potential, $v_{\rm Hxc}\equiv v_{\rm H}+v_{\rm
   xc}$, by inverting Kohn-Sham equations with the full CI densities as an input (Methods).
 In Figure~\ref{potz5} we compare the CI xc-potential (dots in panel a) with 
 the LDA (panel b). We see that for large $a$  (strong-correlation) a
 large barrier develops in the bond while this effect  is missed by the
 LDA.

 Supplementary Figures~\ref{potr4p5} and \ref{potz1} show that the arguments developed rigorously
 for large $Z$ still describe the behavior for fixed $a$ reducing $Z$,
 or for $Z=1$ changing $a$.   Notice, however
 that for the latter  (Fig.~\ref{potz1}) the barrier becomes smaller as one goes to larger
  $a$ which may seem surprising according to our previous arguments. 
 Also in all weakly-correlated cases (Fig.~\ref{potz5} and Supplementary
 Figures \ref{potr4p5} and \ref{potz1}) a broad small barrier appears and LDA
 performs reasonably well. This barrier is of a different
 physical origin as discussed below.

 The solid  lines in Figs.~\ref{potz5}a  (see also Supplementary
 Figures~\ref{potr4p5}a and \ref{potz1}a) represent a reverse-engineering potential (REP) obtained 
 from an approximate analytical inversion of KS equations combined with
 the solution of the lattice problem (L+REP) constructed from
 orthogonalized 1s atomic orbitals (Methods). This inversion,
 involving a Laplacian, is a singular task so that subtle errors in the
 density propagate to give large errors in the potential and the
 task may seem hopeless with such a rough basis. Since a large source of
 error comes from the basis, we first derive equations to 
 optimize it (Methods and
 Refs.~\cite{Spaek2000,Brosco2015}).  Fortunately,  the
 solution of these equations is not needed. Instead, one finds that they
 can be used to eliminate the Laplacian, yielding equations which
 are much less sensitive to the basis and can be evaluated with approximate orbitals (see \ref{sec:insens-reverse-engin}).   
 The resulting expressions for the Hartree-exchange-correlation
 potential read,  
 \begin{equation}\label{vhxcroFinT}
 v_{\rm Hxc}({\bf r})= v_{\rm xc}^{{\rm kin}}({\bf r})+v_{\rm xc}^{{\rm resp}}({\bf r})+v_{\rm Hxc}^{{\rm cond}}({\bf r})
 \end{equation}  
 with
 \begin{eqnarray}
  %\label{eq:kinconresp}
    v_{\rm xc}^{{\rm kin}}({\bf r})&=&\frac{(1-q^2)}{2}
    \frac{|\phi_a({\bf r})\vec\nabla\phi_b({\bf r})-\phi_b({\bf
        r})\vec\nabla\phi_a({\bf r})|^2}{n^2({\bf r})} \label{kin}\\
   v_{\rm xc}^{{\rm resp}}({\bf r})&=&\frac{t (1-q)[\phi_a({\bf
       r})-\phi_b({\bf r})]^2}{n({\bf r})}+\delta \epsilon_g
 \label{resp}\\
 v_{\rm Hxc}^{{\rm cond}}({\bf r})&=& \frac{1}{Z}\int 
 \frac{n_2({\bf r},{\bf r}')}{n({\bf r})|{\bf r}-{\bf r}'|} d{\bf
   r}'. \label{cond}
 \end{eqnarray}
 where  $\delta \epsilon_g>0 $ is a small positive constant
 and $n_2({\bf r},{\bf r}')$ is the two-body density\cite{buijse1989}. 
 The terms correspond one by
 one to the partition of the xc-potential obtained
 by Buijse et al.~\cite{buijse1989}. The present expressions, however, in terms
 of the lattice hopping reduction factor $q$ are new.

 As shown in Fig.~\ref{potz5} and Supplementary Figures~\ref{potr4p5} and \ref{potz1}, using L+REP eqs.~\eqref{kin}-~\eqref{cond} yields a
 very accurate xc-potential from the weakly to the strongly-correlated regime. This
 holds even in the case $Z=1$ where the influence of higher energy
 orbitals could have spoiled the agreement (Supplementary Figures~\ref{potr4p5} and \ref{potz1}).

 The first two terms in equation~\eqref{vhxcroFinT} are order $Z^0$ while the
 last term is at most of order $1/Z$ thus these equations show
 explicitly that the 
 $v_{\rm xc}$ has parts with anomalous coupling constant
 scaling. Specifically the first term, which vanishes in the weakly-correlated limit ($q=1$), 
 yields the Mott barrier.

 In the large $Z$ limit and for large $a$ the above equations become particularly simple for
 the barrier height. Subtracting a small spurious constant term, there is no
 contribution from $v_{\rm xc}^{{\rm  resp}}$ and the height separates in two contributions, one of
 Mott-Hubbard type and one of Coulomb form with different $Z$-scaling, 
 \begin{eqnarray}
 v_{\rm xc}^{{\rm kin}}(0)&\simeq&\frac{(1-q)}{2(1+q)} \label{eq:vheigthkinsim}   \\
 v_{\rm Hxc}^{{\rm cond}}(0)&\simeq& \frac1{Z a}.\label{eq:vheigtcondsim}
 \end{eqnarray}

\begin{figure}
 $$\includegraphics[height=5cm]{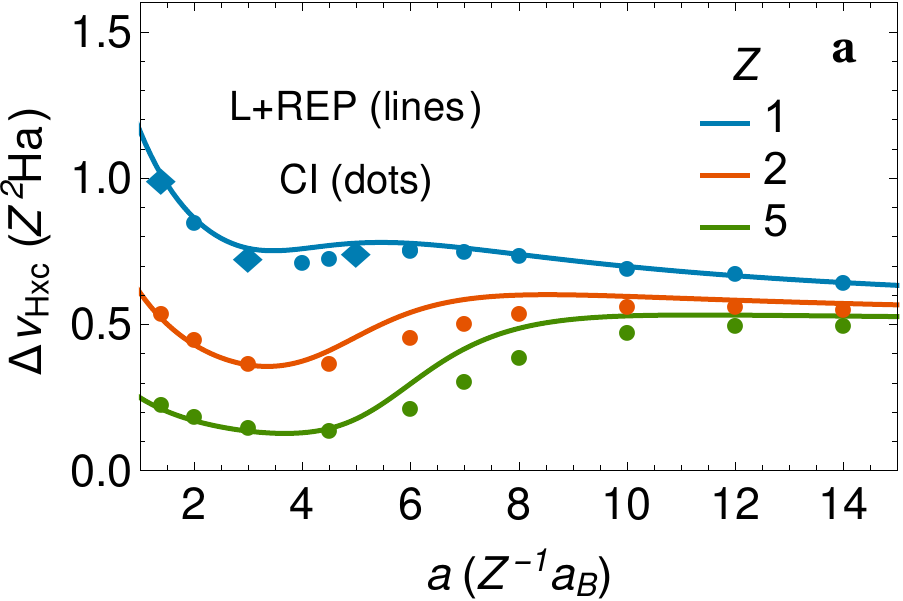}
 \includegraphics[height=5cm]{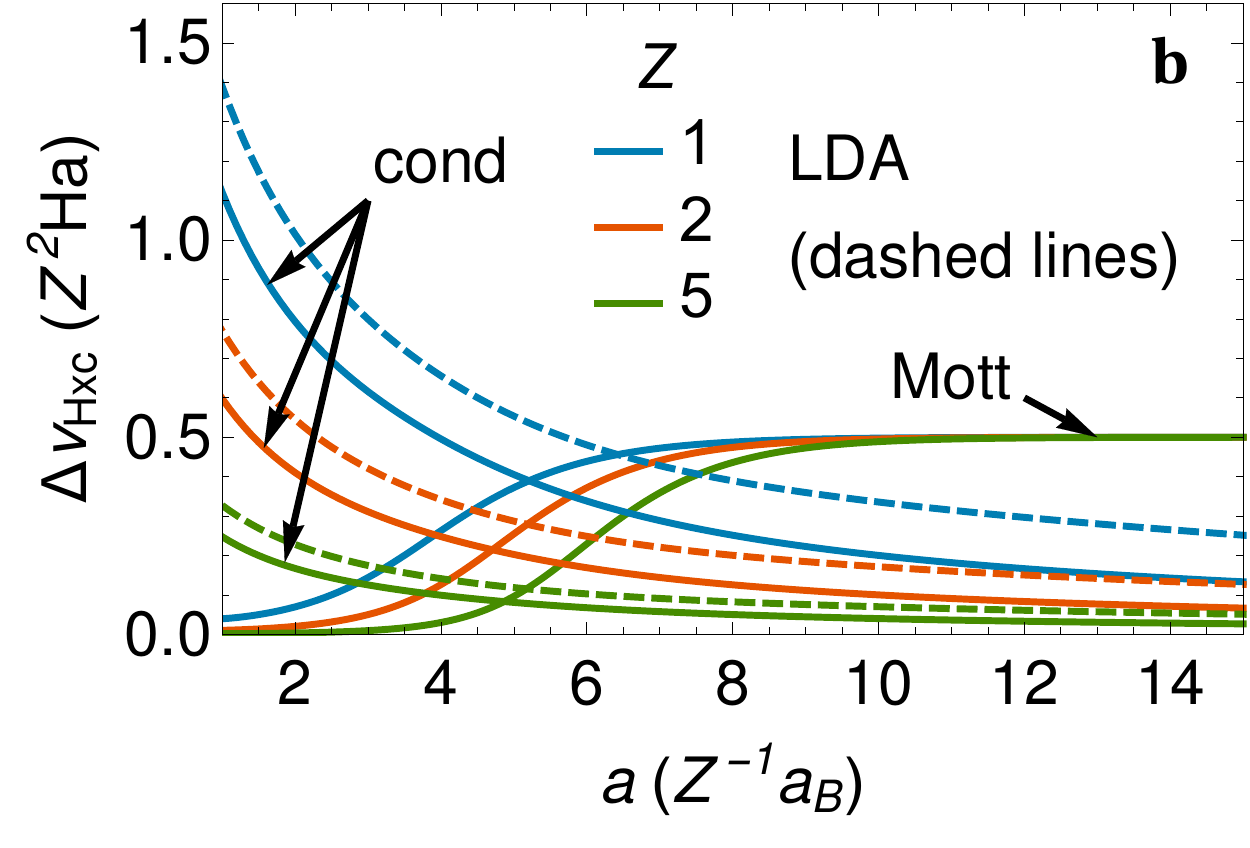}$$
 \caption{
 {\bf  Barrier height.} Height of the Hartree exchange-correlation potential at the
 bond  midpoint for the two-site system  as a function of internuclear
 separation for different $Z$. In {\bf a} full lines are obtained using
 the L+REP approximation subtracting a small spurious term at infinity
 (equations~\eqref{eq:vheigthkin}-\eqref{eq:vheigtcond} of Methods).
 Dots are obtained inverting the full CI charges. For $Z=1$, we
 include also data obtained with an accurate variational wave function
 (Methods). Diamonds are from Ref.~\cite{buijse1989}.  {\bf b} Shows
 separately the anomalously scaling Mott barrier height contribution, equation~\eqref{eq:vheigthkin} 
 and the contribution scaling as $1/(Za)$ (cond) from equation~\eqref{eq:vheigtcond}.  
 The dashed lines are the LDA results. 
 }
 \label{fig:vheight}
  \end{figure}

 Fig.~\ref{fig:vheight}a shows the barrier height as a function of $a$
  for different $Z$. The dots are the numerical data while the full
 lines are obtained from equations~(\ref{eq:vheigthkin})-(\ref{eq:vheigtcond}) of Methods.  Fig.~\ref{fig:vheight}b shows the barrier
 separated in the two components.  LDA yields a
 quite good account of the $Z^{-1}$ component while the $Z^0$ component
 is completely missed. One can also see that the $Z^0$ component just
 reflects the $q$ behavior as a function of distance shown in
 Fig.~\ref{fig:qdu}. For Hydrogen, an accidental compensation of the
 distance dependence of the two
 components explains why the crossover was not identified in the
 potential before.

 \subsection{Generalization to many atoms}

 To illustrate the relevance of these results for extended systems, 
 we introduce a simple generalization of the L+REP equations
 (\ref{kin})-(\ref{cond}) to the  many-atom many-electron
 case. Equation~\eqref{cond} can be used without modification.
 Equations~\eqref{kin},~\eqref{resp} are important only in the strong
  correlation regime. In this case, since electrons are localized, we
  expect that the wave function resembles the two-site wave function. 
 Thus eqs.~\eqref{kin},~\eqref{resp} are generalized by replacing the site
 labels $a$ and $b$ by site index $i$ and $j$ and summing over all  $\la i,j\ra$
  nearest-neighbor sites. $q$ is obtained from the solution of the lattice
  problem in the geometry considered.

 Figure~\ref{fig:rhodr}a shows the barriers for a four site H chain
 while b compares  a quantum chemistry computation of the charge with the KS charge corresponding to the
 L+REP. The small difference between the two charges indicates that the
 L+REP is accurate also in this case.

 \begin{figure}[tb]
 $$\includegraphics[height=5cm]{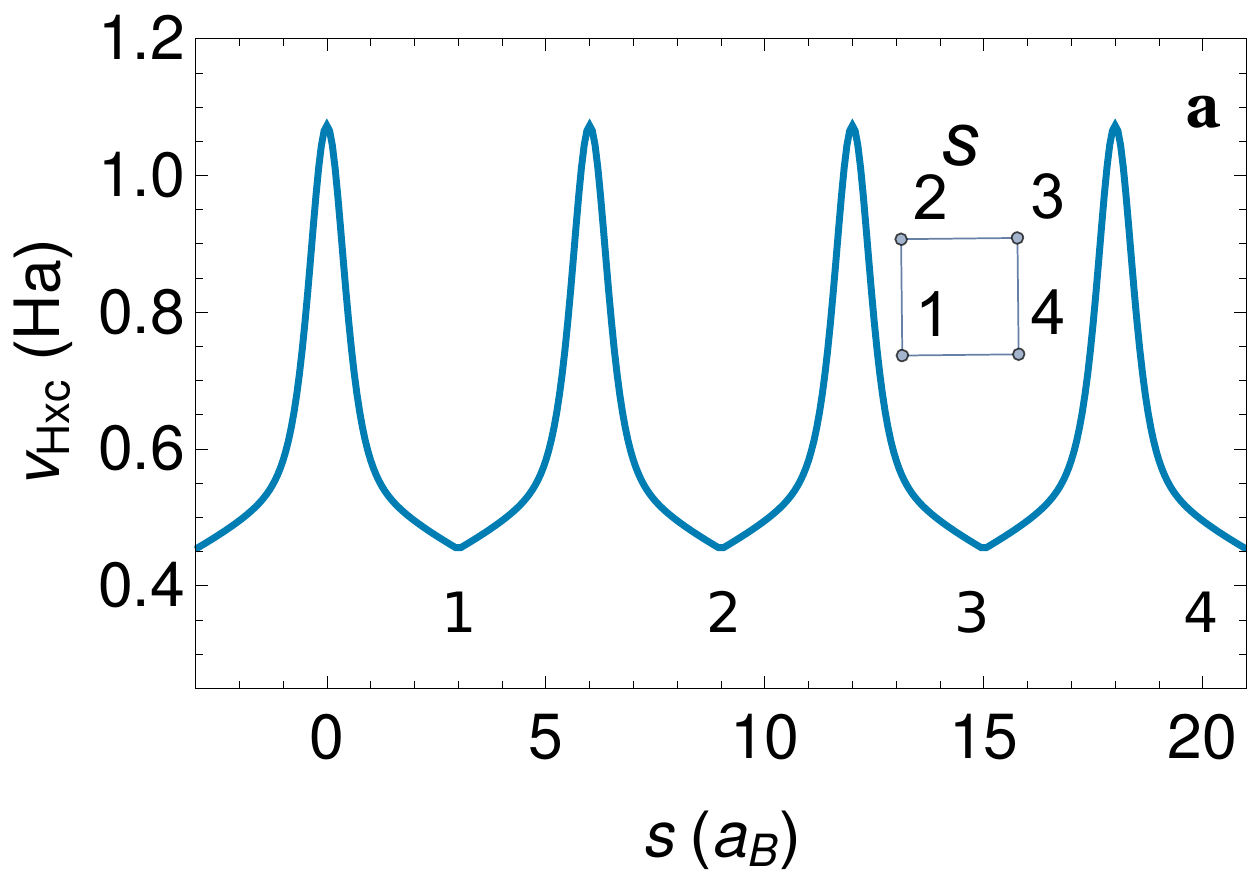}
 \includegraphics[height=5cm]{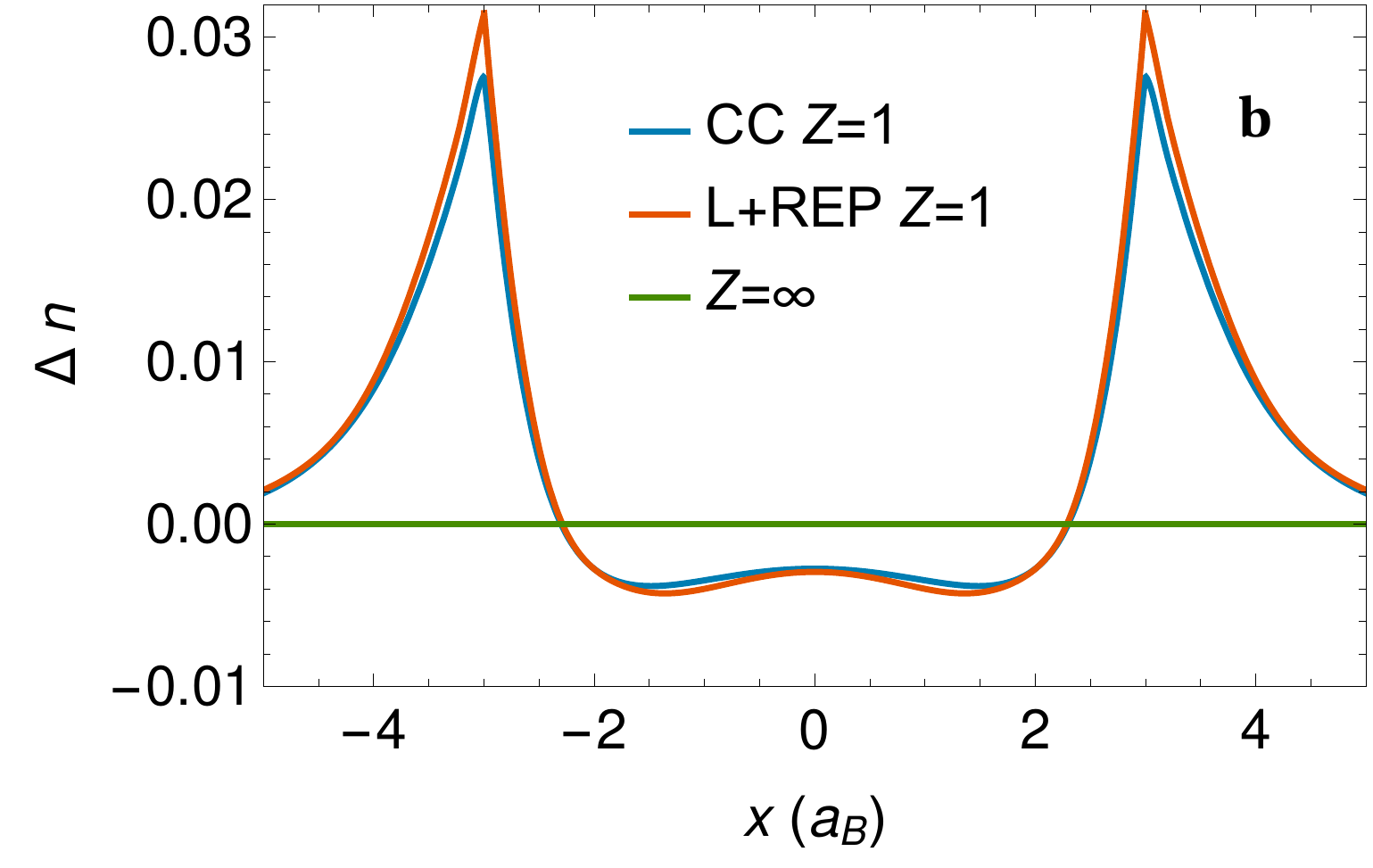}$$
 \caption{{\bf Mott barriers for a four atom chain}. Panel 
 {\bf a} shows $v_{\rm Hxc}$ along the path shown in the inset for 4 H atoms
 arranged in a square with interatomic distance $a=6\  a_B$. Panel 
 {\bf b} compares the interacting charge computed with an accurate
 quantum chemistry method (CC) and the charge from the 
 solution of the  Kohn-Sham potential obtained with the L+REP method. 
 Both are displayed as the difference between the interacting
 and the non-interacting charge ($Z=\infty$).  }\label{fig:mottbar4}  
 \end{figure}

 \section{Discussion}

 It is often assumed that Kohn-Sham DFT bands do not show any
 renormalization due to interactions.   Our results clearly indicate
 that the exact Kohn-Sham DFT bands will show band narrowing due to a suppression of tunneling  stemming from Mott barriers (c.f. Fig.~\ref{potz5}a,
 \ref{fig:mottbar4}a,~\ref{potr4p5}a and \ref{potz1}a). 
 However, we have also shown that as soon as any conventional local 
 semilocal, or hybrid functional is used, this effect is lost.  Thus, although
 the assumption is in principle incorrect, in most practical
 computations available at present it is correct because the Mott barrier effect is not included.

 The bond charge problem studied here provides different insight with respect to the
 fractional charge and fractional spin analysis\cite{cohen2008}, which correspond to the very stretched limit. Here we analyze explicitly a less extreme case, gaining important information on the more challenging intermediate correlation regime.

 We find that the xc-potential separates naturally into two components, one
 with a conventional coupling constant scaling ($1/Z$) and one with an
 anomalous coupling constant 
 scaling $Z^0$. Standard functionals capture only the first
 component, thus, those approaches in which Mott-Hubbard correlations are
 incorporated at a second stage modifying the conventional DFT band structure  
 through DMFT or Gutzwiller approximations can be seen as a way to
 take into account the missing component.\cite{Kotliar2006Electronic,Bunemann2005,Wang2008,Ho2008,Yao2011}
 We hope our results would allow to reformulate this approach in a more
 rigorous way to avoid double counting problems.

  Clearly to obtain the anomalous coupling constant 
  scaling directly from a functional is a highly non-trivial task.
 The present L+REP approach is a  shortcut to this problem. In
 particular equation~\eqref{kin} establishes an intimate relationship
 between the suppression of tunneling in the lattice model and in the
 Kohn-Sham description linking in a neat way the two worlds of lattice models and
 DFT in the continuum.

 \clearpage

 \section*{Methods}

 \subsection{Lattice Model.}
 \label{sec:generalized-hubbard}

 The 1B model is obtained by expanding the field operators
 $\psi_\sigma({\bf r})$ and $\psi^\dag_\sigma({\bf r})$   %which annihilates one electron at point ${\bf r}$ 
 in a 
 minimal basis consisting of  $N$ Wannier orbitals $\phi_i({\bf
   r})\equiv \la{\bf r}|i \ra$ centered on the $N$ sites of the lattice,  
 \be\label{eq:fields}
 \psi_\sigma({\bf r})=\sum_{i\sigma} \phi_i({\bf r})  c_{i\sigma}+...\qquad {\rm and} \qquad \psi^\dag_\sigma({\bf r})=\sum_{i\sigma} \phi^*_i({\bf r})  c^\dag_{i\sigma}+...
 \ee
 where the ellipsis indicate neglected higher energy states.
 For the scaling arguments it is enough to define $\phi_i({\bf  r})$  as Wannier orbitals obtained from the lower band of Bloch states that
 diagonalize the non-interacting Hamiltonian (non-interacting Wannier
 orbitals). In the 1B limit defined in the main article they are very
 similar to orthogonalized atomic 1s orbitals. More accurate 
 Wannier orbitals are discussed below.

 Using eqs. \eqref{eq:fields}, the 1B generalized Hubbard Hamiltonian  corresponding to the continuum model defined in eq.~\eqref{eq:hcont} can be cast as
 \begin{equation}
   \label{eq:latticemodel}
 H_{\rm 1B}=H_h+H_w= \sum_{ ij\sigma}h_{ij}c_{i\sigma}^{\dag}c_{j\sigma}
 +  \frac12\sum_{ijkl\sigma\sigma'}w_{ij,kl} c_{k\sigma}^{\dag}c_{i\sigma'}^{\dag}c_{j\sigma'}c_{l\sigma}
 \end{equation}
 with $h_{ij}=\la\ i|\hat h| j\ra$ and 
 \begin{equation}
   \label{eq:coulint}
 w_{ij,kl}   =   \int d^3{\bf r} d^3{\bf r}'\phi_k^*({\bf r})\phi_i^*({\bf r}') w({\bf r}, {\bf r'})
 \phi_j({\bf r}')\phi_l({\bf r}),
 \end{equation}
 with $w({\bf r},{\bf r'})$ denoting Coulomb interaction:
 \begin{equation} 
 \label{eq:int} w({\bf r},{\bf r'})=\frac{1}{Z|\br-\br'|}. 
 \end{equation}
 $H_{\rm 1B}$ is written in terms of bare matrix elements as,
 for example, the onsite Coulomb interaction $U_0\equiv w_{ii,ii}$.
 The effect of orbitals outside the basis is often accounted for\cite{Kotliar2006Electronic} by replacing the
 bare matrix elements by screened matrix elements. However in the 1B limit
 this effect can be neglected and we drop the not (i.e. $U_0=U$). 

  Equations \eqref{eq:fields} can be also employed to relate the one- and two-particle densities, respectively $n({\bf r})$ and $n_2({\bf r}, {\bf r}')$  
 to the one and two-particle lattice density matrices as follows:
 \begin{equation}
   \label{eq:nvsr}
 n({\bf r})=\sum_\sigma\langle \psi_\sigma^\dagger({\bf r})\psi_\sigma({\bf r}) \rangle=\sum_{ij}  
 \phi^*_j({\bf r})\phi_i({\bf r})\rho_{ij}.
 \end{equation}
 and 
 \begin{equation}
   \label{eq:n2vsr}
 n_2({\bf r},{\bf r'})=\sum_{\sigma\sigma'}\langle  \psi_{\sigma'}^\dagger({\bf r}') \psi_{\sigma}^\dagger({\bf r})\psi_\sigma({\bf r})\psi_{\sigma'}({\bf r}') \rangle=\sum_{ijkl}  
 \phi^*_i({\bf r})\phi_j({\bf r}) \phi^*_k({\bf r}')\phi_l({\bf r}')D_{ij,kl}.
 \end{equation}
 { where we
 defined the spin-integrated two-body lattice density matrix,
 $D_{ij,kl}=\sum_{\sigma\sigma'}\la c_{i\sigma }^{\dag}c_{k\sigma'
 }^{\dag}c_{l\sigma'}c_{j\sigma} \ra$. 
 The density of the Hartree state is recovered by inserting in eq.~\eqref{eq:nvsr} the Hartree lattice density matrix, $\rho_{ij}^H$,  {\sl i.e.}, for two sites $\rho_{ab}^H=1$, while for a chain of atoms,}
 \begin{equation}
   \label{eq:rhoh1d}
  \rho_{ab}^H=\frac2N\sum_{|k a|<\pi/2} \cos{k a}=\frac2\pi.  
 \end{equation}

 { Let us now focus on the  two-site case. Labelling the two sites as $a,b$,  the 
 one-particle lattice Hamiltonian is simply defined by $v=h_{aa}=h_{bb}$ and $-t=h_{ab}$}, while the 
 Coulomb operator reads\cite{hub63,Amadon1996}, 
 \begin{eqnarray}\label{eq:coul}
 H_w&=&U\sum_{i=a,b} n_{i\uparrow }n_{i\downarrow }+V n_a n_b+ t_c \sum_{\sigma} (n_{a\bar\sigma }+n_{b\bar\sigma })( c_{a\sigma }^{\dag}c_{b\sigma }+H.c.)\\
 &+& K\sum_{\sigma\sigma'}  c_{a\sigma }^{\dag}c_{b\sigma' }^{\dag}  c_{a\sigma' }c_{b\sigma }
 + K'\sum_{\sigma}  c_{a\sigma }^{\dag}c_{a\bar\sigma }^{\dag}
 c_{b\bar\sigma }c_{b\sigma }\nonumber
 \end{eqnarray}
 where $V=w_{aa,bb}$ denote the inter-site repulsion, $K=w_{ab,ba}$,
 is the direct exchange interaction, $K'=w_{ab,ab} $ can be thought as a
 Coulomb repulsion among bond-charges, alternatively it can be seen as a
 pair-hopping term. For real orbitals $K=K'$.  Eventually $t_c=w_{aa,ab}$ is a
 correlated hopping term and  it can be considered as the contribution
 of the Hartree potential to the hopping.

 The two-site lattice model can be solved exactly. The ground state
 energy is given by,  
 \begin{eqnarray*}
 E_G &=&\frac 12\left( U+V+K+K^{\prime }-\Delta ^{Gw}\right)  
 \end{eqnarray*}
 where
 \[
 \Delta ^{Gw}=\sqrt{\left( U-V-K+K^{\prime }\right) ^2+16\left(
 t-t_c\right) ^2}.
 \]
 The ground state is
 \begin{eqnarray*}
 \label{wf2site}
 \left| \Psi _{s,0} \right\rangle  &=&\frac{|\Phi_{ \rm HL}\ra
 +\gamma |\Phi_{\rm ion}\ra}{
 \sqrt{1+\gamma ^2}}
 \end{eqnarray*}
 where $|\Phi_{HL}\rangle$ and $|\Phi_{\rm ion}\rangle$ are defined by
  \begin{eqnarray}
   \label{eq:gwfdef}
 |\Phi_{\rm HL}\rangle&=&\frac1{\sqrt{2}}( c_{a\uparrow}^\dagger
 c_{b\downarrow}^\dagger + c_{b\uparrow}^\dagger
 c_{a\downarrow}^\dagger   )  |\emptyset \rangle,\\
 |\Phi_{\rm ion}\rangle&=&\frac1{\sqrt{2}}( c_{a\uparrow}^\dagger
 c_{a\downarrow}^\dagger + c_{b\uparrow}^\dagger
 c_{b\downarrow}^\dagger   )  |\emptyset \rangle
 \end{eqnarray}
  and $\gamma$ is given by
 \begin{equation}
 \gamma =\frac{\Delta ^{Gw}-\left( U-V-K+K^{\prime }\right)
 }{4\left( t-t_c\right) }.
 \label{eq:gamma}
 \end{equation}
 We notice that $\gamma$ coincides with  Gutzwiller variational
 parameter (See \ref{gwf}) since in the present case the Gutzwiller
 wave function equals the exact  ground-state  $\left|
   \Psi_{s,0}\right\rangle$. 
 The exact hopping reduction factor reads,
 \begin{equation}
 \label{eq:gutz}
 q=\frac{2\gamma}{1+\gamma^2}. 
 \end{equation}
 Fig.~\ref{fig:qdu}b was obtained from
 eqs.~\eqref{eq:coulint}~\eqref{eq:gamma}~\eqref{eq:gutz} using
  orthogonalized atomic 1s orbitals.

 \subsection{Optimum basis set.}
 \label{sec:gutzw-wave-funct-opt}

  As mentioned in main text and noted by several authors (see {\sl e.g.}\cite{Gaiduk})
  the inversion of KS equations to determine the KS potential is a difficult
 task.  For example the value of the KS potential with respect to
 the value at infinity (assumed to be zero) is determined by 
 the decay rate of the tail of the density far away from the
 nuclei\cite{Almbladh1985}. Thus an exponentially small error in
 the density coming from an approximate orbital basis 
 can produce an order-one error in the potential.  Thus, we
 first present a computation of the optimum minimal basis set to expand
 the field operator. For simplicity we restrict to a lattice wave-function which depends on the single parameter $\gamma$ but the method can be
 easily generalized to the full set of parameters which specify the lattice wave function.\cite{Brosco2015}

 The variational energy is written as a functional of the Wannier states to be optimised and of the parameters that specify the lattice wave function as follows\cite{Spaek2000,Brosco2015}:
 \begin{eqnarray}
   \label{eq:varener}
 E[\phi_i,\phi_i^*,\gamma]&=&\sum_{ij}h_{ij} \rho_{ji}+\frac12\sum_{ijkl}w_{ij,kl} D_{kl,ij}+\sum_{ij}\epsilon_{ij}(\la\phi_i|\phi_j\ra-\delta_{ij})  
 \end{eqnarray}
 where  $\epsilon_{ij}$ is  an Hermitian matrix of Lagrange parameters that 
 implements the constraint of the orthonormality of the orbitals.
 The variation with respect to $\phi_i^*$ leads to,
 \begin{equation} \label{eq:varEphi-exp}
 \sum_j
 \lf( \rho_{ij} \hat h({\bf r})+\sum_{kl}D_{ij,kl} w_{kl}({\bf r})-\epsilon_{ij}\rg)\phi_j({\bf r})=0
 \end{equation} 
 where we introduced the potential 
 $
  w_{kl}({\bf r})=\int d{\bf r'} \phi_k^*({\bf r}') w({\bf r},{\bf r}')
  \phi_l({\bf r}'). %
 $

 Along with the minimisation with respect to $\gamma$, equations~\eqref{eq:varEphi-exp}
  define a set of closed integro-differential equations. Both problems have to be solved
 self-consistently since the electronic matrix elements in
 equation~\eqref{eq:gamma} depend on the orbitals [equation~\eqref{eq:coulint}]
 which in turn depend on the lattice density matrices through
 equation~\eqref{eq:varEphi-exp}.
  The latter can be further simplified by
 transforming to the  natural orbital basis\cite{Brosco2015} where the one-body density matrix and the  Lagrange multiplier matrix become
 diagonal,  $\bar \rho_{\mu\nu}=\delta_{\mu\nu} \bar \rho_{\mu}$ whith 
 the bar denoting matrix elements in the rotated basis.

 Now we restrict to the two-site case. Minimization respect to $\gamma$
 yields back eq.~\eqref{eq:gamma}. 
 The natural orbitals are,
 \begin{equation}
 \label{eq:nator}
 \psi_0({\bf r})=\frac{\phi_a({\bf r})+\phi_b({\bf r})}{\sqrt{2}},\,\,\quad \psi_1({\bf r})=\frac{\phi_a({\bf r})-\phi_b({\bf r})}{\sqrt{2}}.
 \end{equation}%
  and the density matrix takes the familiar Gutzwiller
  form\cite{met87,Vollhardt1984}, 
  \begin{equation}
    \label{eq:qgw}
 \bar \rho_0=1+q,\quad\quad\bar \rho_1=1-q,
  \end{equation}
 with $q$ given by equation~\eqref{eq:gutz}, 
 while for $D$ we have 
 \be \label{eq:barD} \begin{array}{l}\bar D_{000}=1+q,\,\bar D_{110}=1-q, \,\bar D_{010}=\bar D_{100}=0\\[0.2cm]
 \bar  D_{001}=\bar D_{011}=\bar D_{111}=\bar D_{101}=-\sqrt{(1+q)(1-q)}.\end{array}\ee

 The equations for the states $\psi_{\mu}$ can be cast as effective
 single-particle equations, {\sl i.e.} 
 \begin{eqnarray} \label{opt-eq}
 \left(-\frac{1}{2}\nabla^2+v_{\rm ext}
 +v_{\mu}({\bf r})\right)\psi_\mu=\o_\mu\psi_\mu
 \end{eqnarray}
 where we set $\epsilon_\mu=\rho_\mu\o_\mu$  and the potentials $v_0$
 and $v_1$ are defined by 
  \begin{eqnarray} v_0({\bf r})&=&\bar w_{00}({\bf
      r})-\sqrt{\frac{1-q}{1+q}}\bar w_{01}({\bf r})\frac{\psi_1({\bf r})}{\psi_0({\bf r})},\label{v0}\\
  v_1({\bf r})&=&\bar w_{11}({\bf r})-\sqrt{\frac{1+q}{1-q}}\bar
  w_{10}({\bf r})\frac{\psi_0({\bf r})}{\psi_1({\bf
      r})}.\label{v1}\end{eqnarray}
 Eventually one can solve equations~\eqref{opt-eq}-\eqref{v1} to obtain the
 natural orbitals and invert equations \eqref{eq:nator} to obtain the
 Wannier orbitals.\cite{Brosco2015} In the following we will not follow
 this route but we will use the above expressions to derive a set of
 equations for the xc potential that can be evaluated directly with
 approximate orbitals.

 \subsection{Reverse-Engineering Potential}
 \label{sec:kohn-sham-equations}

 Here we derive the L+REP equations \eqref{kin}-\eqref{cond} for an
 homoatomic bond. 

 For a closed shell system the Kohn-Sham equations read 
 \begin{equation}
   \label{eq:khon-sham-sh}
 \left(-\frac12\nabla^2+v_{KS}({\bf r})-\epsilon_{k}\right) \varphi_{k}({\bf r})=0     
 \end{equation}
 where $v_{KS}$ is given by equation~\eqref{eq:vks}.
 The density is given by,
 \begin{equation}
   \label{eq:nks}
 n({\bf r})=2\sum_{k\in occ} \varphi_{k}^*({\bf r})\varphi_{k}({\bf r}).  
 \end{equation}
 where $k$ labels the Kohn-Sham states and the sum runs over occupied
 states.

 For two-electron systems only the
 $k=0$ state is populated in equation~\eqref{eq:nks} so
 $$\varphi_{0}({\bf r})=\sqrt{\frac{n({\bf r})}2}$$
 and the effective Kohn-Sham potential can be easily expressed as follows up to
 a constant $\epsilon_0$: 
 \begin{equation}\label{eq:vks-2e}
 v_{KS}({\bf r})=\epsilon_0+\frac{\nabla^2 \sqrt{n({\bf r})}}{2\sqrt{n({\bf r})}}.
 \end{equation} 
 The constant can be determined by requiring that the potential at
 infinity is zero which defines $\epsilon_0$ as the highest occupied
 Kohn-Sham orbital eigenvalue.  According
 to DFT Koopmans theorem\cite{levy1984,Almbladh1985} it is related to
 the ionization energy by $\epsilon_0=-E_I$.

 Subtracting the external potential one obtains the
  Hartree-exchange-correlation potential $v_{\rm Hxc}$ [c.f. equation\eqref{eq:vks}],   
  \begin{equation}\label{eq:vhxc-2e}
 v_{\rm Hxc}({\bf r})=v_{KS}({\bf r})-v_{ext}({\bf r}).
 \end{equation} 

 Given two Wannier orbitals, $\phi_a$ and $\phi_b$,
 to expand the lattice model (not
 necessarily optimized) we can define bonding and antibonding orbitals
 as in eq.~\eqref{eq:nator}.   The density of the two-site problem in
 this basis set can be put as, 
 \begin{equation}
   \label{eq:rog}
 n({\bf r})=\sum_{\mu=0,1} \bar\rho_{\mu} \psi_\mu^2({\bf r})
 = (1+q) \psi_0^2({\bf r})+(1-q) \psi_1^2({\bf r}).   
 \end{equation}
 Replacing equation~(\ref{eq:rog}) in equations~(\ref{eq:vks-2e})-(\ref{eq:vhxc-2e}),
 the Hxc-potential $v_{\rm Hxc}$  can be written as the sum of two
 contributions: 
 \begin{equation}\label{vhxcro}
 v_{\rm Hxc}= v_{\rm xc}^{{\rm kin}}+v_{\rm Hxc}^{{\rm rc}}
 \end{equation}  
  where 
  %
  % \begin{eqnarray}
 \begin{eqnarray}
   v_{\rm xc}^{{\rm kin}}&=&\frac{(1-q^2)}{2} \frac{|\psi_1( {\bf r})\vec\nabla\psi_0( {\bf r})-\psi_0( {\bf r})\vec\nabla\psi_1( {\bf r})|^2}{n^2({\bf r})}\label{vhxc1}\\
   \!\!v_{\rm Hxc}^{{\rm rc}}&=&\frac{\sum_{\mu}\, \bar \rho_\mu
     \psi_\mu( {\bf r})\nabla^2\psi_\mu( {\bf r})}{2\,n({\bf r})}-v_{\rm ext}(
   {\bf r})-E_I \label{vhxcr}
 \end{eqnarray}
 Transforming back to atomic orbitals in the first equation one obtains
 equation~\eqref{kin}. Ironically the contribution to $v_{\rm Hxc}$
 which is the hardest to conventional DFT methods, i.e. the part
 scaling as $Z^0$, does not require the use of the orbital optimization
 equations and is already in its final form for numerical evaluation
 with suitable approximate orbitals (we use orthogonalized atomic
 orbitals as discussed in Supplementary Note 3). Furthermore 
 setting $q=0$ one recovers the exact results of Helbig et
 al. (cf. eq.~(11) of Ref.~\cite{Helbig2009Exact}) in the extremely 
 correlated case which are here generalized to arbitrary correlation.

 We find, however, that evaluation of the remaining terms with
 approximate orbitals yields a potential in
 gross disagreement with numerical methods due to the presence of the 
 Laplacian in eq.~\eqref{vhxcr} (see
 \ref{sec:insens-reverse-engin}). 
 Thus, in the following we assume that the orbitals are
 optimized. Surprisingly this condition can be relaxed in the final
 equations effectively eliminating the strong sensitivity to the basis.

 Inserting the optimisation equations, eqs.~(\ref{v0})-(\ref{v1}), in equation~(\ref{vhxcr}) we can cancel the external potential term and eliminate the Laplacian to obtain, 
  \begin{equation}\label{vhxc2-1}
 v_{\rm Hxc}^{{\rm rc}}({\bf r})=-\frac{\sum_{\mu} \bar \rho_{\mu}[\o_\mu-v_\mu({\bf r})]\psi_{\mu}^2({\bf r})}{n({\bf r})}-E_I.
 \end{equation} 
 This expression can be transformed to a more transparent and computationally more convenient form by splitting $v_{\rm Hxc}^{{\rm rc}}$ in two parts:
 \begin{equation}\label{passage1}
 v^{{\rm rc}}_{\rm Hxc}({\bf r})=v_{\rm xc}^{\rm resp}( {\bf r})+v_{\rm Hxc}^{\rm cond}( {\bf r})
 \end{equation} 
 where 
  \begin{equation}v_{\rm Hxc}^{\rm cond}( {\bf r})=\frac{\sum_{\mu}\bar \rho_\mu\psi^2_\mu( {\bf r}) v_\mu( {\bf r})}{n({\bf r})}.
 \end{equation} 
 and
 \begin{equation}
   \label{eq:resp}
 v_{\rm xc}^{\rm resp}( {\bf r})=-\frac{(\o_1-\o_0)\bar \rho_1\psi_1( {\bf r})^2 }{n({\bf r})}+E_G-\epsilon_g-\o_0.  
 \end{equation}
 { In deriving the above equations, we used the definition of the density, eq.~\eqref{eq:rog}, and we set 
 $E_I=\epsilon_g-E_G $ with $\epsilon_g$ denoting  the one particle ground state energy. }

  Using the explicit expression of the potentials $v_\mu( {\bf r})$
  [equations~(\ref{v0})-(\ref{v1})] we can eventually recast the ``cond'' term as  
 \begin{equation}v_{\rm Hxc}^{\rm cond}( {\bf r})=\frac{\sum_{\mu\nu}\sqrt{\bar \rho_\mu\bar \rho_\nu}\psi_\mu\psi_\nu \bar w_{\mu\nu}}{n({\bf r})}.\label{vcond1}
 \end{equation} 
 By a direct calculation one can then easily recover the expression of  $v_{\rm
   Hxc}^{\rm cond}$ first obtained by Bujise et al.  by a completely different
 method\cite{buijse1989}, namely equation~\eqref{cond} of main text, with the  two-particle density defined as usual as: 
 $n_2({\bf r},{\bf r}')=\sum_{\sigma,\sigma'}
 \langle  \psi_\sigma({\bf r})^{\dag} \psi_{\sigma'}({\bf r'})^{\dag} 
  \psi_{\sigma'}({\bf r'})  \psi_\sigma({\bf r})\rangle$.

 To arrive at the final expression for $v_{\rm xc}^{\rm resp}$, given in eq.~\eqref{resp} we use the two following
 identities:
  \begin{eqnarray}
 & &\o_1-\o_0=\bar h_{00}-\bar h_{11}\equiv-2 t,\label{Dst}\\
 & &   E_G=\omega_0+\bar h_{00}=2\omega_0-\la\psi_0|v_0|\psi_0\ra.\label{koop}
   \end{eqnarray}
 { Before coming to the proof of the above identities, let us  note that equation \eqref{Dst}  implies that the
 optimized bonding orbital $|\psi_0\rangle$ corresponds to the highest
 Lagrange multiplier, {\sl i.e.} $\o_0>\o_1$, the opposite of the naive guess. This
 sign change is fundamental to
 obtain the correct { behaviour of $v_{\rm xc}^{\rm resp}$} and the correct decay of the density. 
 The relation $\o_0>\o_1$ indeed implies that the behaviour of the density at large distances is governed by $\o_0$
 that, as suggested by \eqref{koop}, is correctly related to the ionisation energy of the system.\cite{Almbladh1985}
 Note however, that within our approximations, $\o_0$ differs from the ionisation energy by a small constant, $\d\epsilon_g=h_{00}-\epsilon_g$. 
 The latter is due to the relaxation of the orbitals upon ionisation and it tends to zero in the large $Z$ limit.
 By replacing eqs. (\ref{Dst})-(\ref{koop}) in equation~\eqref{eq:resp} 
 we arrive at the final expression for $v_{\rm xc}^{\rm resp}$,
 eq.~\eqref{resp},  and we can easily show  that 
 $\lim_{r\rightarrow\infty}v_{\rm xc}^{\rm resp}({\bf r})=\d\epsilon_g$

 \noindent Equation~\eqref{Dst} can be proved by noting that from equations~\eqref{opt-eq} it follows that
 \begin{equation} \o_1-\o_0=\bar h_{11}-\bar h_{00}+\la \psi_1 |v_1|\psi_1\ra-\la\psi_0| v_0|\psi_0\ra\label{eqepsilon}\end{equation} 
 while from the definition of $q$ and of the $v_\mu$ we have that:
 \begin{eqnarray}
 \la \psi_1 |v_1|\psi_1\ra\!-\!\la\psi_0| v_0|\psi_0\ra\!&=&\!\bar w_{11,11}\!-\!\bar w_{00,00}-\frac{4 \gamma\, \bar w_{01,01}}{1+\gamma^2}\nonumber\\
 &=& 2 (\bar h_{00}-\bar h_{11})\label{h11h00}
 \end{eqnarray}
 which replaced in equation~\eqref{eqepsilon} leads to equation \eqref{Dst}. 
 Notice that in the last step on the r.h.s. of equation \eqref{h11h00} we have used the explicit expression of $\gamma$ in terms of one and two-electron integrals given in equation~\eqref{eq:gamma}.\\
  In order to demonstrate equation~\eqref{koop} we can start  from the ground state energy %, given in equation~\eqref{Eg}
 which can be recast in terms of the energies %
 $\omega_i$
 as follows
 \begin{equation} E_G=\frac{1+q}{2}\lf(\o_0+\bar h_{00}\rg) +\frac{1-q}{2}\lf(\o_1+\bar h_{11}\rg)  \end{equation}  
 which using equation~\eqref{Dst} in turn leads to
 \begin{equation} E_G=\o_0+\bar h_{00}=\o_1+\bar h_{11}\end{equation} 
 which concludes the proof of eq.~\eqref{koop}.

 To obtain the L+REP results shown in the figures we replaced the
 bonding and antibonding states by appropriate linear combinations of  atomic orbitals, {\sl i.e.}
 we set
 \be \psi_0({\bf r})=\frac{\varphi_{1s}({\bf r}-{\bf R}_a)+\varphi_{1s}({\bf r}-{\bf R}_b)}{\sqrt{2(1+ S)}}\qquad \psi_1({\bf r})=\frac{\varphi_{1s}({\bf r}-{\bf R}_a)-\varphi_{1s}({\bf r}-{\bf R}_b)}{\sqrt{2(1- S)}}\ee
 where $\varphi_{1s}({\bf r})=e^{-\xi |{\bf r}|}\sqrt{\xi^3/\pi}$, 
 $S$ denotes the overlap integral between $\varphi_{1s}({\bf r}-{\bf
   R}_a)$ and $\varphi_{1s}({\bf r}-{\bf R}_b)$ and $\xi$ was obtained variationally.}

 In this way we obtain
 \begin{eqnarray}
 v_{\rm xc}^{{\rm kin}}(0)&=&\frac{(1-q)}{2(1+q)}\frac{1+S}{1-S} \xi^2 \label{eq:vheigthkin}   \\
 v_{\rm Hxc}^{{\rm cond}}(0)&=& \frac{1}{2Z(1+S)}\int \frac{|\varphi_{1s}({\bf r}-{\bf R}_a)+\varphi_{1s}({\bf r}-{\bf R}_b)|^2}{|{\bf r}|} d{\bf r}.\label{eq:vheigtcond}
   \end{eqnarray}
 The integrals in $v_{\rm Hxc}^{{\rm cond}}$ are
 known\cite{Slater1965}. For large $Z$ (1B limit) the crossover regime between
 strong and weak correlation lays at large separation $a$.  Thus we can
 set $S\simeq 0$ and $\xi=1$ yielding
 equations~\eqref{eq:vheigthkinsim}-\eqref{eq:vheigtcondsim}. 

 \subsection{Quantum chemistry computations}
 Accurate densities in Fig.~\ref{fig:rhodr} where  obtained using 
 full CI with the ORCA computer code.\cite{orca} Very large basis sets were used in order to have well converged densities. In particular we used the fully uncontracted aug-mcc-pV8Z\cite{Peterson} for the case $Z=1$ and the same basis set with the exponent appropriately scaled for the systems with $Z >1$.

 Accurate Kohn-Sham potentials used as reference in
 Figures:~\ref{potz5}, \ref{fig:vheight}, \ref{potr4p5} and \ref{potz1}
 were extracted by using equation \eqref{eq:vks-2e} starting from the
 full CI densities obtained as above. The density was computed on a
 cubic grid and  spurious features in the KS potential  due to the
 basis set  were removed by applying the scheme described in
 Ref.~\cite{Gaiduk}.
 For the four site case (panel b of Fig~\ref{fig:mottbar4})  full CI is
 not feasible. Therefore we used the Coupled Cluster method using
 aug-cc-pV5Z basis set with the code Orca~\cite{orca}.

 The LDA potentials, and non-interacting densities were calculated using cp2k code.\cite{cp2k}

 \section*{Acknowledgments}

 This work was supported by the Italian Institute of
 Technology through the project NEWDFESCM. PG-G acknowledges financial
 support from the European Research Council under H2020/ERC Consolidator Grant “corr-DFT” (Grant No. 648932).

\section*{Supplementary Figures}

\setcounter{figure}{0}
\makeatletter 
\renewcommand{\thefigure}{S\@arabic\c@figure}
\makeatother

\begin{figure}[h!]
$$\includegraphics[height=5cm]{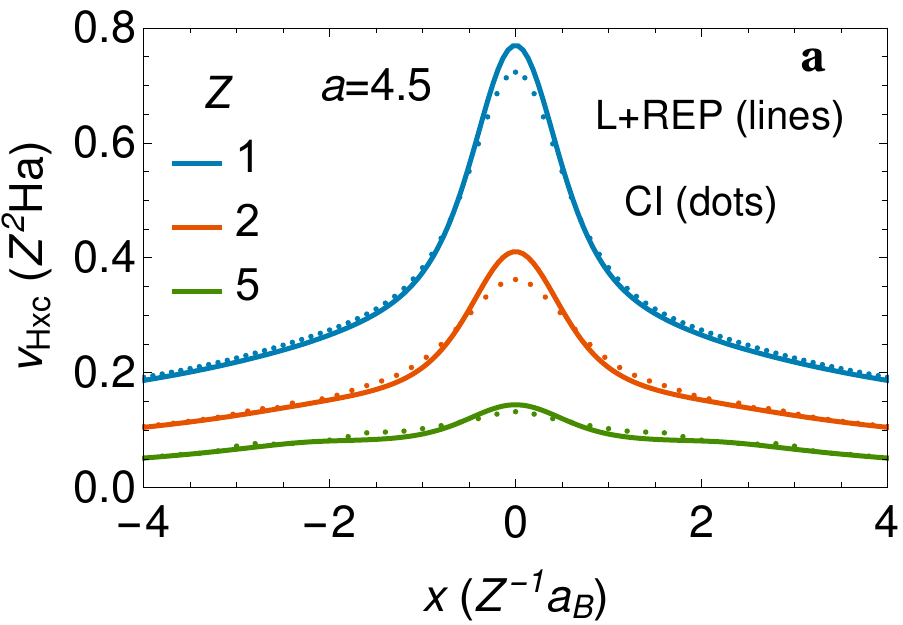}
\includegraphics[height=5cm]{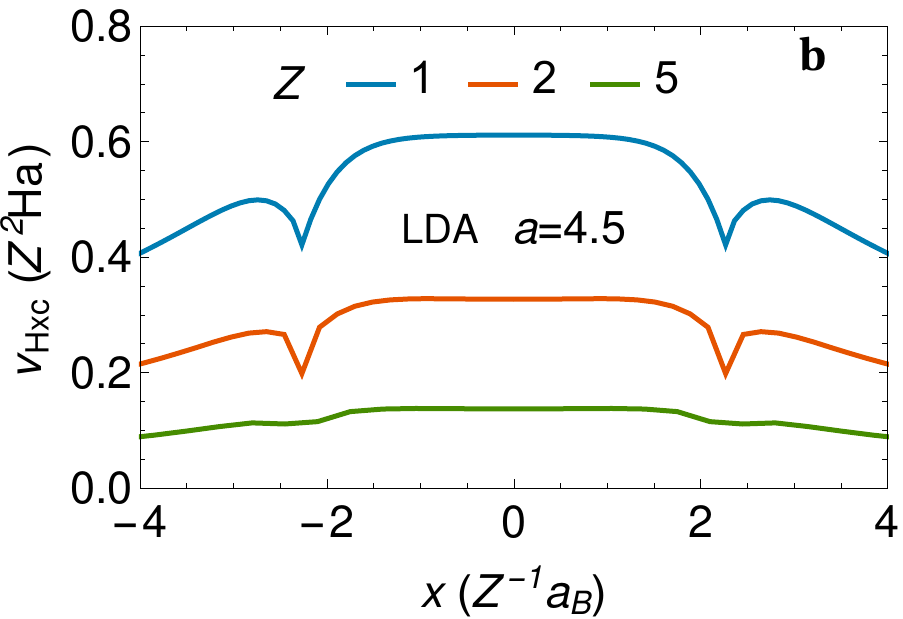}$$
\caption{{\bf Mott Barrier vs.} $Z$.
Hartree exchange-correlation potential for $a=4.5 Z^{-1}a_{B}$ and different
  values of $Z$  for the two-site
  system  as a function of $x$ for $z=y=0$ obtained inverting the full CI 
ground state density  as explained  in Methods (dots in {\bf a})  
and in the LDA\cite{cp2k} ({\bf b}). The lines in {\bf a} are analytic expressions based on the
L+REP approximation.  The
 strength of the correlations for each case is defined by the value of $q$ in Fig.~\ref{fig:qdu}.
 }
\label{potr4p5}
 \end{figure}

\begin{figure}[h!]
$$\includegraphics[height=5cm]{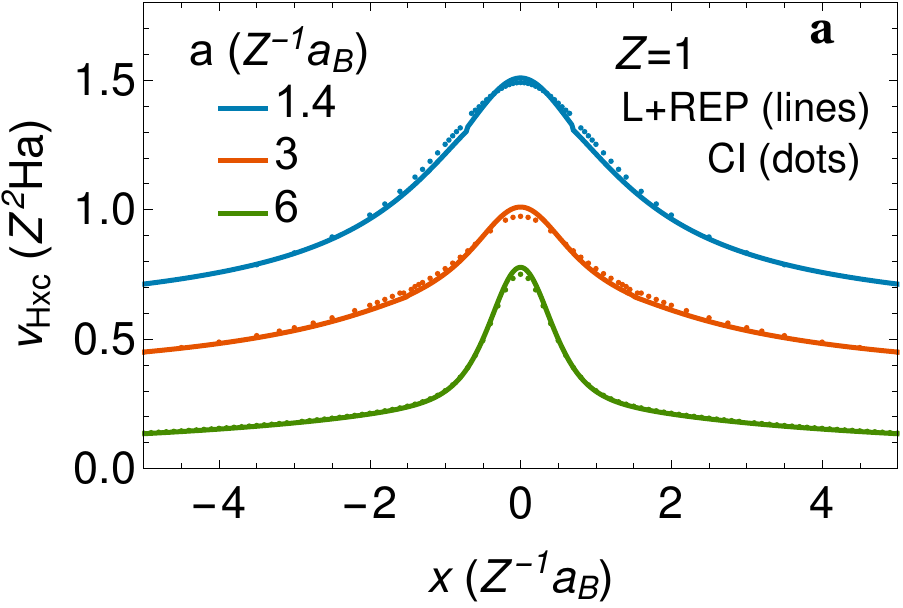}
\includegraphics[height=5cm]{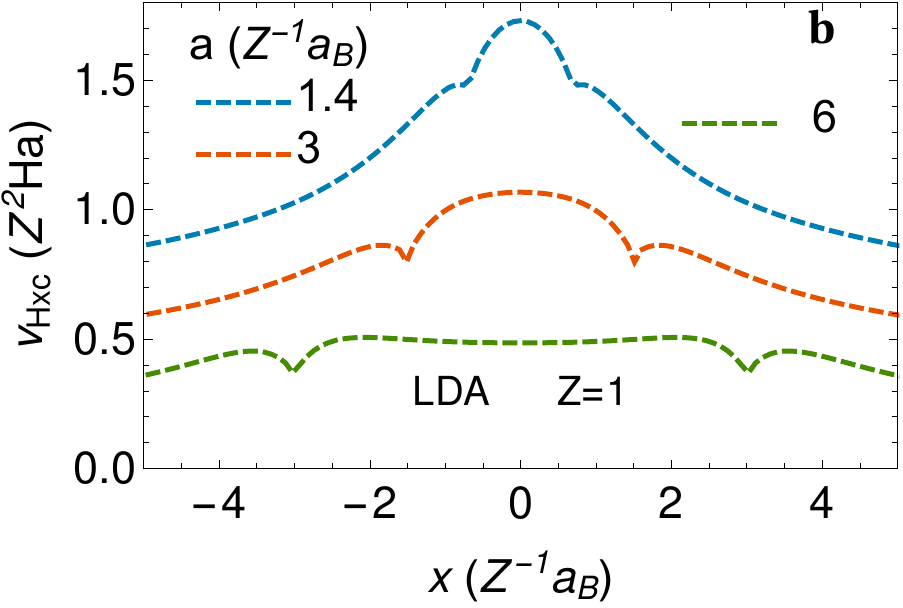}$$
\caption{{\bf Hartree-exchange-correlation potential for the stretched
  Hydrogen molecule.} We show $v_{\rm Hxc}$ as a function of $x$ for
$z=y=0$ for different values
  of $R$. The dots in {\bf a} where obtained inverting the full CI 
ground state density  as explained  in Methods while the 
 lines are the L+REP results.  Panel {\bf b} shows the
LDA  results. The cases $a=1.4$ and $a=3$ where shifted by 0.5 and 0.25
 respectively ($Z^{2}$Ha) for clarity. The first two cases are in the
 weakly and intermediate correlation regimes while the case $a=6$ is
 in the strong coupling regime (c.f. Figs.~\ref{fig:qdu}{\bf b} and
 \ref{fig:vheight}{\bf b}).  
}
\label{potz1}
 \end{figure}

\clearpage

\setcounter{section}{0}
\makeatletter 
\renewcommand{\thesection}{Suppplementary Note~\@arabic\c@section}
\makeatother

\section{Gutzwiller Wave Function}
\label{gwf}
The correlation induced changes in bond charges, described by the
parameter $q$  are rooted in the competition between tunneling energy 
and Coulomb induced localization. The Gutzwiller wave function
is a simple tool to study such competition.

\noindent For a lattice of identical atoms, the Gutzwiller wave function can be
written as,\cite{gut63,Gutzwiller2009Gutzwiller,Vollhardt1984,met87,geb90} 
\begin{equation}
|\Psi_\gamma \rangle=\frac{\gamma^D}{C^{1/2}_\gamma}|\Psi_0 \rangle \label{eq:psigamma}\end{equation} 
where $|\Psi_0 \rangle$ is a Slater determinant, $D=
\sum_i n_{i\uparrow }n_{i\downarrow }$ counts the total double
occupancy, $\gamma$ is a variational parameter and $C_\gamma=\langle
\Psi_0 |\gamma^{2D}|\Psi_0 \rangle$ a normalization constant. 
For $U>0$ the operator $\gamma^D$ decreases the
weight of configurations with double occupied sites.  
The Gutzwiller variational problem  can be solved exactly in
infinity and in one-dimension\cite{geb90,met88}. 
In the two-site case the Gutzwiller wave function coincides with
the exact expression given in eq.~\eqref{wf2site} of the main
article, $|\Psi_\gamma \rangle=\left| \Psi _{s,0} \right\rangle$,  
and interpolates between the Hartree-Fock (HF) and the Heitler-London (HL) solutions  recovered respectively for $\gamma=1$ and $\gamma=0$.%

Unlike the two-atom case the Gutzwiller wave function does not yield the exact
solution of the infinite-dimensional problem. However comparing with
the exact solution, which can be obtained numerically using dynamical
mean-field theory, one finds that it gives a remarkably accurate description of the metallic
phase. In the insulating phase it yields $q=0$, in contrast with the 
finite value of  $q$ of  the exact solution (c.f.
Fig.~\ref{fig:qdu}a). Furthermore the exact solution shows a
Mott transition at the kink position in the blue curve of
Figure~\ref{fig:qdu}  in main text while in the Gutzwiller approximation the
transition occurs at  $U=U_{BR}$ with $U_{BR}$ denoting the Brinkman-Rice $U$.\cite{Gutzwiller2009Gutzwiller}

\section{ Testing  functionals}
\label{sec:testing-functionals}

The exchange correlation potential of local and semilocal functionals
can be written as, 
\begin{equation}
  \label{eq:vxcslco}
v_{\rm xc}=v_{\rm xc}(a_B^3\tilde n, a_B^4\tilde\nabla \tilde n, a_B^5\tilde\nabla^2
\tilde n,... )=v_{\rm xc}(Z^3 n, Z^4 \nabla  n, Z^5\nabla^2  n,... )   
\end{equation}
where quantities with tilde have usual dimensions and quantities
without tilde are rendered dimensionless with the rescaled units defined in
the main text, i.e. $\tilde n=Z^3 n /a_B^3$, $\tilde \nabla=Z
\nabla/a_B $, etc.

In the following,  for simplicity, we take the same two-site setting
of the main text. In order to test a functional one should check its
scaling properties in the 1B limit, thus we should
take the limit of $Z\rightarrow \infty$ keeping $a\sim \log(Z)$ such
that $t/U$ is constant. More precisely, for large $Z$ atomic orbitals become
asymptotically correct and matrix elements take the values\cite{Slater1965} $U=5/(8Z)$ and
$t=e^{-a}$ (Ha$Z^2$). So the 1B limit condition can be written as, 
\begin{equation}
  \label{eq:1b}
 e^{-a}=\frac{5t}{ 8Z U }.    
\end{equation}

It is convenient to interrogate functionals at the origin where the
part of the potential scaling as $Z^0$ should be larger.
As an initial guess we can use the correlated density eq.~\eqref{eq:rog}
taking again atomic orbitals to expand the wave-function, 
\begin{equation}
  \label{eq:n0}
n(0)=(1+q)\frac{2}\pi e^{-a}= (1+q)\frac{2}\pi \frac{5t}{ 8Z U }.
\end{equation}
As $Z$ is driven to infinity the atoms run away from the origin and
the rescaled density is driven to zero as $1/Z$.  On the other hand,
the physical density $\tilde n$ grows as $Z^2$ so one needs the
high-density limit of $v_{\rm xc}$ in eq.~\eqref{eq:vxcslco}.  

For the LDA gradient terms are not present and $v_{\rm xc}$ behaves as, \cite{PerWan-PRB-92}
\begin{eqnarray}
  \label{eq:vlda}
v_{\rm xc}^{\rm LDA}&=& A a_B \tilde n^{1/3} {\rm Ha}\nonumber \\
&=& A  \frac1Z  n^{1/3} \sim \frac1{Z^{4/3}} \ \ \ \ (Z^2 {\rm Ha}), 
\end{eqnarray}
where $A$ is a constant and the second line shows that in rescaled
units $v_{\rm xc}^{\rm LDA}\rightarrow 0$. Imposing selfconsistency
does not change this result since the density is driven to the
non-interacting density which is given by eq.~\eqref{eq:n0} with
$q=1$.  One can also consider points which are at a fixed position 
respect of the nucleus in rescaled units. In this case the rescaled
density remains asymptotically constant as $Z$ grows and $v_{\rm
  Hxc}^{\rm LDA}\sim 1/Z$. 
Therefore, since there is not term scaling as $Z^0$, we conclude, 
that LDA can not describe Mott phenomena in the 1B limit.

For other functionals  specific forms must be considered as the final
result will depend on the scaling properties of the potential in the
limit in which all arguments of eq.~\eqref{eq:vxcslco} become
large. However, in general the behavior will be dominated by the
high-density limit of the functional where most conventional
functionals converge to the non-interacting limit and have no portions scaling as $Z^0$ in rescaled units. 
As an example we follow Van Leeuwen and Baerends\cite{Leeuwen1994} and
we evaluate the potential at the origin for a specific form of the Generalized Gradient Approximation (GGA),
\begin{eqnarray}
  \label{eq:vgga}
v_{\rm xc}^{\rm GGA}&=&B a_B \frac{\nabla^2 \tilde n}{\tilde n^{4/3}} {\rm Ha}\nonumber \\
&=&B  \frac1Z \frac{\nabla^2  n}{ n^{4/3}}=B  \frac1Z \frac{1}{ n(0)^{1/3}}=B  \left(\frac{4\pi U/t}{5(1+q)}\right)^{1/3} \frac1{Z^{2/3}}    \ \ \ \ (Z^2 {\rm Ha}).  
\end{eqnarray}
Fixing $t/U$ and letting $Z\rightarrow \infty$, one sees that the
potential goes to zero more slowly than LDA but still not enough to
provide the barrier.  Again, selfconsistency does not improve the result.

Finally, hybrid functionals do not solve the problem either as they
incorporate a portion of exact exchange which also collapses to zero
in the high-density limit.

\section{Weak sensitivity of the Reverse
  Engineering Potential to Orbital Basis Errors}
\label{sec:insens-reverse-engin}

Inverting KS equations usually requires an extremely accurate basis to
expand the wave-functions and the correlated densities. 
Such large sensitivity of the potential is rooted in
 the presence of the Laplacian appearing in our case in
 eq.~\eqref{vhxcr} leading to the ``cond'' and ``resp''
 contributions while for the ``kin'' contribution the problem does not
 arise.   To  exemplify the problem, lets $\psi_\mu$ be
molecular orbitals (MO) satisfying eq.~\eqref{opt-eq} with $v_{\mu}$
set to zero and 
$\omega_0=\epsilon_g$. Using these definitions we can rewrite eq.~\eqref{vhxcr} as, 
\begin{equation}
 v_{\rm Hxc}^{rc}= -\frac{t (1-q) 2 \psi_1^2({\bf r})}{n({\bf r})}+E_I-\epsilon_g.\ \ \ \ (\rm MO)
\label{vhxcmo}
\end{equation} 
where all quantities should be evaluated in the MO basis. 
By comparing the above expression with equations \eqref{resp} and \eqref{cond}
given in main text, we notice that:
(i) there is no ``cond'' contribution scaling as $1/(Z\,R)$;
(ii) apart from constants eq.~\eqref{vhxcmo} resembles
$v_{\rm xc}^{\rm resp}$ but it has the opposite sign. A different but
still wrong result is obtained using in eq.~\eqref{vhxcr}  an
approximated Gutzwiller density constructed using linear combinations
of atomic orbitals.  Equations~\eqref{kin}-\eqref{cond} solve this extreme
sensitivity problem thanks to the elimination of the Laplacian using a
saddle condition for the energy. Indeed, contrary to
eq.~\eqref{vhxcmo}, they reproduce one by one with the
equations of Ref.~\cite{buijse1989} simply evaluating their
expressions in the same minimal basis we use. 

Clearly,  even 
using eqs.~\eqref{kin}-\eqref{cond} errors in the basis still
will lead to errors in the potential  however they can be shown to be
quite mild in comparison with the errors which are incurred using
equations containing the Laplacian.  Indeed  $v_{\rm xc}^{{\rm
    kin}}({\bf r})$,  vanishes in the weak-coupling regime
($q\rightarrow 1$) while in the strong coupling it converges  to the
exact result in the Heitler-London limit ($q=0$) obtained by Helbig et
al.~\cite{Helbig2009Exact}. In the intermediate coupling regime it 
provides a smooth interpolation between these two extremes with the
crossover at the right place in parameter space ensuring that errors
remain small.
  
As regards, $v_{\rm xc}^{{\rm resp}}({\bf r})$ we instead see that it
is  the sum of two parts:  a spurious constant $\d\epsilon_g$ (which
by definition does not contribute to the barrier height) and a
position dependent potential which  also does not contribute to the
barrier height since it vanishes both at the origin and at infinity.  The
latter contribution is always quite small for homoatomic systems since
the prefactor $(1-q)t$  is small both in the weak and in the strong
correlation limit. Notice that $\d\epsilon_g$ would cancel in the
approximation in which one uses the  same orbitals to expand the
two-site Hubbard model in the one- and two-atom cases.

The last contribution,  $v_{\rm Hxc}^{{\rm cond}}({\bf
  r})$ can be quite large for small $Z$ and decays slow with
distance so it is critical to get it accurately.   The two-particle density can be evaluated 
expanding the field operators in the minimal basis and using the lattice ground state 
to evaluate the 2-body lattice density matrix. Fortunately, since there are no
derivatives affecting the orbitals (and some are integrated)  an
extremely accurate knowledge of the orbitals is not required to calculate  $v_{\rm  Hxc}^{{\rm cond}}({\bf r})$.

We conclude saying that, while in principle $\phi_a$ and $\phi_b$ in
equations~\eqref{kin}-\eqref{cond} of the main text are optimized
Wannier orbitals whose shape minimize the total energy they can
be safely replaced by approximate orbitals.  
In doing the plots, they were approximated with linear combinations of
atomic 1s-orbitals as defined in the following note.

\section{Orthogonalized atomic orbitals}

As explained in the text we approximate the optimized orbitals $\psi_{\mu}$ with linear combination of atomic orbitals as follows:
\begin{eqnarray}\label{eq:ortog}
\psi_0 &\simeq&\frac{\varphi _{1s}({\bf r}-{\bf R}_a)+\varphi_{1s}({\bf r}-{\bf R}_b)}{\sqrt{
2(1+S)}}, \\
\psi_1 &\simeq&\frac{\varphi _{1s}({\bf r}-{\bf R}_a)-\varphi_{1s}({\bf r}-{\bf R}_b)}{\sqrt{2(1-S)}}.
\end{eqnarray}
with $\varphi_{1s}({\bf r})=e^{-\xi |{\bf r}|}\sqrt{\xi^3/\pi}$ and  $S$ denoting the overlap integral between $\varphi_{1s}({\bf r}-{\bf R}_a)$ and $\varphi_{1s}({\bf r}-{\bf R}_b)$
Accordingly the optimised Wannier orbitals are approximated as: 
\bea 
\phi_a&\simeq&\lf(\varphi _{1s}({\bf r}-{\bf R}_a)-\alpha\, \varphi_{1s}({\bf r}-{\bf R}_b)\rg)/\sqrt{N}\\
\phi_b&\simeq&\lf(\varphi _{1s}({\bf r}-{\bf R}_b)-\alpha\, \varphi_{1s}({\bf r}-{\bf R}_a)\rg)/\sqrt{N}
\eea
where $\a=(1-\sqrt{1-S^2})/S$ and $N=(1-S^2)(1+\alpha^2)$.
Also the one and two-centers integrals defining the parameters of the generalised Hubbard model were calculated using the above approximate expressions for $\phi_a$ and $\phi_b$.
Within this approximation the parameters $t$, $U$, $V$, $t_c$ and $K$
are then simply linear combination of the same parameters calculated with atomic orbitals which are exactly known\cite{Slater1965} and in the equations below are indicated with a tilde. We have,
\begin{eqnarray*}
t &=&\frac{\tilde t}{1-S^2}\\
U &=&\frac 1{2(1-S^2)^2}\left[ \left( 2-S^2\right)
\tilde{U}-4\tilde{t}
_cS+S^2\tilde{V}+2S^2\tilde{K}\right] , \\
t_c &=&\frac 1{2(1-S^2)^2}\left[ 2\left( 1+S^2\right)
\tilde{t}_c-S\left(
\tilde{U}+\tilde{V}+2\tilde{K}\right) \right] , \\
V &=&\frac 1{2(1-S^2)^2}\left[
-4S\tilde{t}_c+2\tilde{V}+S^2\left(
\tilde{U}-\tilde{V}+2\tilde{K}\right) \right] , \\
K &=&\frac 1{2(1-S^2)^2}\left[
-4S\tilde{t}_c+2\tilde{K}+S^2\left(
\tilde{U}+\tilde{V}\right) \right] ,
\end{eqnarray*}
where we took $\tilde{K}'=\tilde{K}$. 

\section{Reverse Engineering Potential for Many-Sites}
\label{sec:reverse-engin-potent}

Analogously to the theory of superexchange\cite{Anderson1950Antiferromagnetism},  in the many-site case, we assume that in the strongly-correlated limit on each bond the exchange-correlation potential has a structure which strongly resembles the one found in the diatomic molecule.
  As we did for the two-electron molecule,  in the many-site many-electron case  we therefore have to: define an appropriate reference lattice model,  find an appropriate single-particle basis set and, after having estimated all parameters of the lattice model calculate the one- and two- body densities.
  
As regards to the first task we start from the single-band generalised lattice model introduced in eq.~\eqref{eq:latticemodel} and, consistently with our assumptions concerning the potential, we 
truncate it neglecting three-site  and interbond correlations. In this way  we essentially replicate the diatomic molecule Hamiltonian  for each couple of sites present in the system and we obtain:
\begin{eqnarray}\label{eq:hchain}
& & H_{1B}\simeq H_{\rm chain}\equiv U\sum_{i} n_{i\uparrow }n_{i\downarrow }-\varepsilon\sum_{i} n_{i} +\sum_{(i, j)}H^{\rm bond}_{i,j}
\eea
 where the sum over $(i,j)$ is not limited to nearest-neighbouring sites but include also next-to-nearest-neighbours and we  set
 \bea
 & & H^{\rm bond}_{ij}=-t( c_{i\sigma }^{\dag}c_{j\sigma }+H.c.)+Vn_{i}n_{j}+t_{c} \sum_{\sigma} (n_{i\bar\sigma }+n_{j\bar\sigma })( c_{i\sigma }^{\dag}c_{j\sigma }+H.c.)+\nonumber\\& & + K\sum_{\sigma\sigma'}  c_{i\sigma }^{\dag}c_{j\sigma' }^{\dag}  c_{i\sigma' }c_{j\sigma }
+ K'\sum_{\sigma}  c_{i\sigma }^{\dag}c_{i\bar\sigma }^{\dag}
c_{j\bar\sigma }c_{j\sigma }\nonumber
\end{eqnarray}
In the above equation, the dependence of the matrix elements, $t$, $V$, $t_c$, $K$ and $K'$ on the distance $i-j$ is implied.

As regards the Wannier states, defining the optimum single-particle basis, we approximate them  as linear combination of atomic orbitals, similarly to what we did in the diatomic case. Namely, we assume that the Wannier orbital corresponding to site $i$ equals a linear combination of atomic orbitals centred on the site $i$ and on nearest- and next-nearest neighbouring sites and it can be written as follows,
\be
\phi_i=\lf(\varphi_i-\alpha\,\varphi_{i+u}-\beta\,\varphi_{i+v}\rg)/\sqrt{N}
\ee
where $u$ and $v$ implement  translations respectively on nearest- and next-nearest neighbouring sites and the weights $\alpha$ and $\beta$ and the normalization $N$ are determined imposing orthonormality.

Using the above equation, as in the diatomic case, we express all integrals appearing in the potentials and the different parameters appearing in the lattice Hamiltonian as linear combinations of  known two-center integrals involving Slater orbitals and  we calculate them  analytically.  
Once this is done, we determine the ground-state of  $H_{\rm chain}$ by exact diagonalization and   we calculate the hopping reduction factor $q$ and the two-body density matrix on the lattice 
needed to finally  estimate the potential by simply generalizing 
eqs.~(\ref{kin})-(\ref{cond}) as explained in the main text.

\end{document}